\documentclass[twoside]{article}
\usepackage{amsmath,amssymb,psfig}

\newtheorem{lemma}{Lemma}
\newtheorem{theorem}{Theorem}

\def\DD{\displaystyle}

\def\Mf{\vphantom{\DD\frac{}{\DD 0}}}

\def\RR{{\mathchoice{{\bf R}}{{\bf R}}{{\rm R}}{{\rm R}}}}

\newcommand{\be}{\begin{equation}}
\newcommand{\ee}{\end{equation}}
\newcommand{\ba}{\begin{array}}
\newcommand{\ea}{\end{array}}

\newcommand{\al}{\alpha}
\newcommand{\bt}{\beta}
\newcommand{\ga}{\gamma}
\newcommand{\de}{\delta}

\newcommand{\la}{\lambda}

\newcommand{\ch}{\chi}

\newcommand{\Ph}{\Phi}
\newcommand{\si}{\sigma}

\newcommand{\De}{\Delta}

\newcommand{\cA}{{\cal A}}
\newcommand{\cB}{{\cal B}}
\newcommand{\cC}{{\cal C}}
\newcommand{\cE}{{\cal E}}
\newcommand{\cL}{{\cal L}}
\newcommand{\Tr}{{\rm Tr}}
\newcommand{\Str}{{\rm Str}}
\newcommand{\Sdet}{{\rm Sdet}}

\let\epsilon=\varepsilon
\let\phi=\varphi

\begin{document}

\medskip
\title{Random Matrices and the Anderson Model}
\author{M. Disertori\\
Theoretische Physik, ETH Z\"urich\\
CH-8093 Z\"urich, Switzerland\\
\\
V. Rivasseau\\
Laboratoire de Physique Th\'eorique\\
B\^atiment 210, Universit\'e Paris XI\\
91405 Orsay Cedex, France}
\maketitle
\begin{abstract}
In recent years, constructive field techniques and the method of
renormalization group around extended singularities have been applied 
to the weak coupling regime of the Anderson Model. It has allowed
to clarify the relationship between this model and the theory
of random matrices. We review this situation and the current program
to analyze in detail the density of states and Green's functions of
this model using the supersymmetric formalism.
\end{abstract}

\section{Introduction}

This small review is devoted to the elementary theory of random matrices
and to the link between this theory and the Anderson model 
of localization/diffusion of a quantum particle in a random
potential.

More precisely we recall first the basic result of random matrix theory,
namely the Wigner's semi-circle law for density of states, and give
its rigorous supersymmetric derivation.

Then we review the Anderson model, introducing the phase space
approach to this model pioneered by Gilles Poirot, and summarizing the results
of \cite{BMR}.

Finally in the last part we  propose some generalizations  of the flip random matrix model of \cite{BMR}
which are closer to the real Anderson model, using in particular some hierarchical approximations.
The control of such more realistic random matrices models  
is an important step towards rigorous theorems about the 
Anderson model in the weak potential phase.

We thank Abdelmalek Abdesselam, Jean Bellissard,
Jacques Magnen and Gilles Poirot for collaboration and discussions on 
various aspects of this work. V. Rivasseau also thanks F. Klopp 
for an invitation to lecture on this subject at Paris XIII- Villetanneuse
University at the origin of this review.

\section{Random Matrices and Wigner's law} 

\subsection{The GUE}

The simplest ensemble of random matrices is the Gaussian unitary ensemble.
It is a probability measure  on random hermitian $N \times N$ complex matrices.
Each coefficient in the upper triangle of the matrix is 
identically independently Gaussian distributed.
Here the matrix is $H=H_{ij} $, $H=H^*$, hence $H_{ij}=\bar H_{ji}$, and
\begin{equation}
P(H) = \frac{1}{Z} \exp(-\frac{N}{2}{\rm Tr} H^* H)
\end{equation} 
$Z$ being a normalization factor. The matrix $H$ is 
made therefore of $N(N-1)/2$ complex variables $H_{ij}$ with $i<j$
and $N$ real ones $H_{ii}$, so there are $N^2$ real random variables in $H$.

Since 
\begin{equation}
Tr H^* H = \sum_{i=1}^N \sum_{j=1}^N |H_{ij}|^2 = \sum_{i} H_{ii}^2 + 2 \sum_{i<j} (\Re
H_{ij}^2 + \Im H_{ij}^2 )
\end{equation} 
we have 
\begin{equation}
Z = 2^{N/2} (\pi/N)^{N^2 /2}
\end{equation} 
and the covariance rule is 
\begin{equation}
<  H_{ij} H^*_{i'j'}> = \frac{1}{N} \delta_{ii'} \delta_{jj'} .
\end{equation} 
The scaling factor $\frac{1}{N}$ has been chosen to keep the typical eigenvalues
of $H$ of size $O(1)$ as $N\to \infty$; indeed the typical size of the eigenvalues of
a random matrix with covariance 1 is obviously of order $\sqrt N$, by the law of large numbers.

Physicists would like to know the statistics of the eigenvalues of $H$, and they are
particularly interested in the two first moments of their distribution,
called the density of states and the two-level correlation function.

The density of states, $\nu(E)$ for a Hermitian matrix
$H$ is the quantity which, when integrated from $-\infty$ to $A$, counts 
the number of eigenvalues of $H$ which are lower or equal to $A$. 
Since $H$ has exactly $N$ real eigenvalues $\lambda_1 ,..., \lambda_N$,
we have 
\begin{equation}
\nu (E) = \frac{1}{N} {\rm Tr} \delta(E-H)
\end{equation} 
so that $\int_{-\infty}^{+\infty} \nu(E) dE =1$.
We can use the standard formula for the Dirac distribution
\begin{equation}
\delta(x-a) = -\frac{1}{\pi} \lim_{\epsilon \to 0_+}  \Im \frac{1}{x-a + i \epsilon}.
\end{equation} 
Hence 
\begin{equation}
\nu (E) = -\frac{1}{\pi N} \lim_{\epsilon \to 0_+}  \Im {\rm Tr} \frac{1}{E-H + i \epsilon}.
\end{equation} 
Physicists call $(E-H \pm i \epsilon)^{-1}$ respectively the retarded and advanced
Green's functions for the Hamiltonian $H$. 

The averaged density of states $< \nu (E)>$ is therefore
\begin{equation}
< \nu (E) > = -\lim_{\epsilon \to 0_+}  \int P(H) dH \frac{1}{\pi N}  \Im {\rm Tr} \frac{1}{E-H + i \epsilon},
\end{equation} 
and $<\nu(E)> dE$ clearly represents the probability for an eigenvalue of $H$ to lie
between $E$ and $E+dE$, with normalization condition $\int < \nu (E)> dE =1$. 

The main results on the GUE ensemble is Wigner's semi-circle law:
\begin{equation}
\lim_{N \to \infty} <\nu (E)> = \frac{\chi_{|E| \le 2}}{\pi} \sqrt{1-E^2 /4} \ .
\end{equation} 
The corresponding curve is really a semi-ellipse, but of course could be changed
into a circle through a slight reparametrization of the covariance of $H$. The normalization
taken here corresponds to $\int E^2 \nu(E) dE =1$.

Wigner's law is a central result. It has been called the non-commutative
analog of the Gaussian law of large numbers \cite{Voi}, and has been proved 
to hold in much more general 
cases than the GUE, for instance for band random matrices \cite{DPS}. 

The next quantity of interest is the 2-level correlation, which allows to know the conditional
probability to find an eigenvalue of $H$ near $E$ knowing already that one eigenvalue
sits at $E$. More precisely it gives the probability to have two eigenvalues separated
by an interval of width $\omega$ centered at $E$, and is therefore
\begin{equation}
R_2  (\omega) = \frac{<\nu(E-\omega/2) \nu(E+\omega/2)>}{<\nu(E)>^2}.
\end{equation} 
In the GUE, eigenvalues are not independent but
tend to "repel" each other. 
This is seen in the following behavior of the 2-level correlation $R_2$ 
\begin{equation}
R_2 (s) = \delta(s) + 1 - \frac{\sin^2 \pi s}{ (\pi s)^2}\ ,  
\end{equation} 
where $s = \omega/\Delta$, and $\Delta$ is the mean level spacing $\Delta = 1/N<\nu (E)>$.
The delta function simply expresses the constraint of presence of an eigenvalue
at $E$. Independence of the eigenvalues
would mean $\lim_{s\to 0}R_2 (s) - \delta(s) =1 $, hence no change in the probability
for a second value to sit near $E$ if a first is present. But here we have 
$\lim_{s\to 0}R_2 (s) - \delta(s) =0 $ because of the $\frac{\sin^2 \pi s}{ (\pi s)^2}  $
term. Hence there is 0 chance for a second eigenvalue to sit near $E$ if a first
one sits at $E$. This is the phenomenon of "eigenvalue repulsion".

Physicists got intuition of this repulsion by
the simple observation of the Vandermonde determinant that appears 
in the Jacobian of the transformation from the initial coefficients of the matrix
to the diagonal eigenvalues and the unitary diagonalizing matrix. In rough terms,
we can diagonalize an Hermitian matrix $H$ through a unitary matrix $U$:
\begin{equation}
H = U \Lambda U^*\quad , \quad U^* = U^{-1} .
\end{equation} 
Then one can write the initial measure $P(H)dH$ in terms of the coefficients
of $\Lambda$ and $U$. Clearly the measure on the unitary group
must factorize from the eigenvalues measure since $P(H)$ is invariant 
through action of the unitary group. Let us explain by a simple argument the well known result
\begin{equation}
P(H) dH = d\mu (U)    e^{-\frac{N}{2}   \sum_{i=1}^N  \lambda_i^2}  \prod_{i< j} (\lambda_i -\lambda_j)^2
\prod_{i=1}^N d\lambda_i   \ .
\end{equation} 
To understand the appearance of the non-trivial Vandermonde factor 
$ \prod_{i< j} (\lambda_i -\lambda_j)^2$ (in addition to the
ordinary trivial factor $P(H)= e^{-\frac{N}{2}   \sum_{i=1}^N  \lambda_i^2}  $)
we need only to compute the Jacobian at origin from the $H$ variables
to the $\la $ variables and the variables parameterizing $U$ near the origin. 
For this purpose, we can derive the relation $U^* U =1$
with respect to a set of local parameters $U_r$ for a local chart of the unitary group 
near the origin. This gives
\begin{equation}
S_{r} = U^* \frac{\partial U}{\partial U_r} = - S^*_r \ ,
\end{equation} 
the tangent space to the unitary group at the origin being the anti-hermitian
matrices. From $H=U\Lambda U^*$ one finds
\begin{equation}
\frac{\partial H}{\partial U_{r}} = \frac{\partial U}{\partial U_r} \Lambda U^* + U \Lambda
\frac{\partial U^*}{\partial U_r}\ ,
\end{equation} 
hence
\begin{equation}
(U^*  \frac{\partial H}{\partial U_r} U)_{ij} = (S_r \Lambda - \Lambda S_r)_{ij} = (S_r)_{ij} (\lambda_j -\lambda _i) .
\end{equation} 
Furthermore 
\begin{equation}
U^* \frac{\partial H_{ij}}{\partial \lambda_k} U = \frac{\partial \Lambda_{ij}}{\partial \lambda_k} 
 =\delta_{ij} \delta_{ik} ,
\end{equation} 
so that the Jacobian to compute is
\begin{eqnarray}
J =  \left|   \ba{cc}  \frac{\partial H_{ii}}{\partial \lambda_k} &   \frac{\partial H_{ij}}{\partial \lambda_k}  \\
 \frac{\partial H_{ii}}{\partial U_r} &  \frac{\partial H_{ij}} {\partial U_r} \ea
 \right| =  \left|   \ba{cc}  1 &  0 \\
 f(\lambda, U) & (\lambda_j - \lambda_i)  S_r (U)  \ea  \right| = g(U) \prod_{i<j} (\lambda_i - \lambda_j) ^2 .
\end{eqnarray} 

Clearly the presence of this Vandermonde determinant means that the eigenvalues of a random
matrix in the GUE case are not independent, but repel each other since the measure
vanish at coinciding eigenvalues. Physically this level repulsion is analogous
to some kind of Pauli exclusion principle between eigenvalues, or to some two body
logarithmic interaction:
\begin{equation}
e^{-\frac{N}{2}  \sum_{i=1}^N  \lambda_i^2}  \prod_{i< j} (\lambda_i -\lambda_j)^2
\prod_{i=1}^N d\lambda_i = e^{-\frac{N}{2}   \sum_{i=1}^N  \lambda_i^2  + 2 \sum_{i<j}  
\log  | \lambda_i - \lambda_j | }  \prod_{i=1}^N d\lambda_i , \label{Vander}
\end{equation} 
which is analogous to Coulomb repulsion in two dimensions (also logarithmic).

It is possible to use the theory of orthogonal polynomials
to analyze the large $N$ limit and recover Wigner's law for this
system or for more complicated non-Gaussian measures on $H$
(for the GUE, orthogonal polynomials are simply Hermite
polynomials).  This is e.g. done in \cite{Mehta}. See also \cite{Past}
for another reference book on the subject.


In this lecture we prefer to stress the supersymmetric 
approach to this problem. It makes particularly transparent
how Wigner's law results from a mean-field theory and a saddle point 
expansion which expresses the subtle balance between the Gaussian and Vandermonde
terms in \ref{Vander}.

\subsection{Supermathematics}

In this presentation we follow the excellent concise review by Mirlin \cite{Mir}.

Grassmann or anticommuting or Fermionic variables are pairs
of independent variables, which for convenience are
noted as complex conjugates, $\ch_1 ,..., \ch_N $, $\ch^*_1,..., \ch_N^*$ with the following properties
\begin{equation}
\ch_i \ch_j = - \ch_j \ch_i \  , \  \ch_i^* \ch_j = - \ch_j \ch_i^* \ , \  \ch_i^* \ch_j^* = 
- \ch_j^* \ch_i^* \   \ ,   
\end{equation} 
\begin{equation}
\int d\ch_i = \int d\ch^*_i  = 0\ , \  \int  \ch_i d\ch_i  = \int  \ch_i^* d\ch_i^*    = \frac{1}{\sqrt{2\pi}} .
\label{grass}
\end{equation} 
Conjugation for Fermionic variables is not involutive but antiinvolutive,
and it does not reverse the ordering of a product: :
\be  (\chi ^*)^* = - \chi  \ , \   (\chi \psi)^* = \chi ^* \psi ^* \ .
\ee
Then for any $N \times N$ matrix $M$ we find using these rules that
\begin{equation}
\int \prod_i d\ch_i^* d\ch_i  e^{- \ch^* M \ch}   = \det (\frac{M}{2\pi}) ,
\end{equation}  
where $\ch$ is now considered a vector $(\ch_1 ,..., \ch_N)$.
For ordinary, also called {\it bosonic}, complex conjugate variables $S, S^*$ we 
would have got instead, requiring $\Re M >0$ for convergence:
\begin{equation}
\int \prod_i dS_i^* dS_i  e^{- S^* M S}   = \det\nolimits^{-1} (\frac{M}{2\pi}) .
\end{equation}  
{\it Supersymmetric} expressions are obtained by assembling bosonic and Fermionic
variables in a symmetric way. For instance a supervector is 
\begin{equation}
\Ph = (S_1 , ...., S_N, \ch_1 , ... \ch_N) \ , \  \Ph^* = (S^*_1 , ...., S^*_N, \ch^*_1 , ... \ch^*_N) .
\end{equation} 
A supermatrix is $M = \left(  \ba{cc}   a & \sigma \\  \rho & b \ea \right)$, $2N $ by $2N$, in which
$a$ and $b$ are ordinary bosonic and $\si, \rho$ are anticommuting variables.
We use Latin indices from 1 to $N$ and Greek indices with values $b$ and $f$ to distinguish
the bosonic and Fermionic parts, so that we write
\begin{equation}
M = M_{\al \beta, ij} \ , \  M_{bb,ij} = a_{ij}  \ , \ M_{ff,ij} = b _{ij} \ , \ M_{bf,ij} = \sigma_{ij}
\ , \ M_{fb, ij}  = \rho_{ij} \ .
\end{equation} 

Traces and determinants generalize into supertraces and superdeterminants:
\begin{equation}
\Str M = \Tr a  - \Tr b  \quad , \quad   \Sdet M = \det (a - \sigma b^{-1} \rho) . \det b^{-1} .
\end{equation} 
We still have
\begin{equation}
\Str \log M = \log \Sdet M \ .
\end{equation} 
Furthermore for any supermatrix (requiring $\Re M_{bb} >0$ for convergence):
\begin{equation}
\int d \Ph^* d \Ph e^{- \Ph^* M \Ph}  = \Sdet M^{-1}  \ .
\end{equation} 
Remark that for a supersymmetric supermatrix $M$, i.e. one in which $\rho = \sigma = 0$ and $a=b$,
this integral is normalized, namely $\Sdet M =1$.

The inverse of a supermatrix can be computed as
\begin{equation}
M^{-1}  = \left(  \ba{cc}   (a -\si b^{-1} \rho)^{-1}   &   -   (a -\si b^{-1} \rho)^{-1} \si b^{-1} \\ 
- b^{-1} \rho  (a -\si b^{-1} \rho)^{-1}    &  b^{-1}  ( 1 + \rho    (a -\si b^{-1} \rho)^{-1} \sigma b^{-1}  )
\ea \right) .
\end{equation} 

Like for an ordinary matrix we can define a resolvent, or two point function
\begin{equation}
\int  d \Ph^* d \Ph \  \Ph_{\al  i}   \Ph^*_{ \beta j}  \; e^{- \Ph^* M \Ph}  = ( M^{-1} )^{\al \beta}_{ij}\  \Sdet M^{-1}  \ .
\end{equation}
 
Further references on supercalculus, supermanifolds can be found in \cite{Zirn}.

\subsection{Wigner's law}

Let us return to the proof of the Wigner's law using the supersymmetric formalism.
The advantage of supersymmetry is to allow the integrated resolvent to be written
as a functional integral in an ordinary field theory because the corresponding  
normalizing determinant is 1. More precisely let us return to
\begin{equation}
< \nu (E) > = - \lim_{\epsilon \to 0_+}  \int P(H) dH \frac{1}{\pi N}  \Im {\rm Tr} \frac{1}{E-H + i \epsilon} \ .
\end{equation} 
We would have indeed with ordinary variables
\begin{equation}
<(E-H+i \epsilon )_{ij}^{-1}>   =  -i \int P(H) dH   
 \frac{\int S^*_i S_j e^{i S^*  (E-H+i\epsilon) S}  dS^* dS}{ \int e^{i S^*  (E-H+i\epsilon) S}  dS^* dS }
\end{equation} 
where the imaginary sign has been chosen so that the bosonic part of the integral 
converges, using the small positive part $e^{-\epsilon  S^* S}$.
The integral of such a quotient is not easy to manipulate. But with supervariables
taking the self-normalization into account we can write
\begin{equation}
(E-H +i\epsilon)_{ij}^{-1}   = -i    
 \int S^*_i S_j e^{ i  \Phi^*  (E-H+i\epsilon)  \Phi }  d\Phi^* d\Phi 
\end{equation} 
so that integration over $H$ can be performed explicitly. Indeed
\begin{equation}
<(E-H +i\epsilon)_{ij}^{-1}>   = -i \int P(H) dH    
 \int S^*_i S_j e^{ i \Ph^*  (E-H+i\epsilon) \Ph}  d\Ph^* d\Ph
\end{equation} 
and the Gaussian integral of an exponential linear in $H$ can be performed exactly,
giving rise to an exponential quartic in $\Ph$, something which physicists
call a "$\Phi ^4$ interaction ":
\begin{equation}
<e^{i \sum_{ij} \Ph_i^* H_{ij} \Ph_j}>  = \exp \bigl\{  - (1/2N) \sum_{ij} (\Ph_i^* \Ph_j) (\Ph_j^*  \Ph_i)           \bigr\} .
\end{equation} 
The essential step is now to recast this quartic term in the form of a so-called "vector model",
that is to factorize the quartic term as a square. Taking carefully into account the anticommutation rules
one finds:
\begin{equation}
\sum_{ij}  (\Ph_i^* \Ph_j) (\Ph_j^*  \Ph_i)    =( \sum_{i}  S_i^* S_i)^2 + 2 ( \sum_{i}  S_i \ch_i^*)( \sum_{i}  S_i^* \ch_i)- ( \sum_{i}  \ch_i^* \ch_i )^2  .
\end{equation} 
When rewritten this way, one can introduce a representation of the quartic
term as an integral over a single supermatrix field $R$. Physicist call this idea
a  "Hubbard-Stratonovich" transformation. In practice the many variables $H$
have been reduced to four "mean field" variables, the coefficients of $R$. We see
that the key fact under this phenomenon is the independence of 
the variables $H_{ij}$\footnote{For a random matrix whose coefficients are not independent,
this mean-field  "Hubbard-Stratonovich" transformation is not valid and Wigner's
law may no longer hold; however if a random matrix has for instance
$N^2/p(N)$ independent coefficients, each spanning an orbit of $p(N)$ sites in the matrix,
we may search for a reduced representation with only $O(p(N))$  "Hubbard-Stratonovich" 
variables and still expect Wigner's law for instance if $p(N)/N \to 0$ as $N \to \infty$, see below.}.

More precisely we have
\begin{equation}
\exp \bigl\{  -  \frac{1}{2N} \sum_{ij} (\Ph_i^* \Ph_j) (\Ph_j^*  \Ph_i)           \bigr\}
= \int dR \exp \bigl\{ -   \frac{N}{2} \Str R^2   -i \sum_{i}  \Ph_i^*  R \Ph_i   \bigr\}
\end{equation} 
with 
\begin{equation}
R =  \left(  \ba{cc}   a & \rho^* \\  \rho & ib \ea \right).
\end{equation} 
Indeed due to the $i$ factor in the fermion-fermion $b$ term, the bosonic part of the quadratic
form $\Str R^2 =  a^2 + b^2 + 2\rho^* \rho $ is positively definite, ensuring convergence.

We can now interchange the $\Ph$ and $R$ integrations and perform the $\Ph$ integrations
which are now quadratic:
\begin{eqnarray}
<\Tr (E-H +i\epsilon))^{-1}> &=& -i  \int dR 
\int d\Ph^* d\Ph \sum_{k} S_k^*S_k  
\nonumber\\ && \hskip-.5cm \exp \bigl\{  i(E+i\epsilon) \sum_{i}  \Ph_i^*   \Ph_i 
-  \frac{N}{2} \Str R^2   -i \sum_{i}  \Ph_i^*  R \Ph_i   \bigr\}
\nonumber\\
 &=&  N \int dR 
(E-R+i\epsilon)_{bb}^{-1} \nonumber\\ &&
\exp \bigl\{-  \frac{N}{2} \Str R^2   -N \Str \log (E-R +i \epsilon )   \bigr\} \, ,
\end{eqnarray}  
so that
\begin{eqnarray}
< \nu (E) > &=& -\lim_{\epsilon \to 0_+}   \frac{1}{\pi }  \Im  \int dR 
(E-R+i\epsilon)_{bb}^{-1}   \nonumber\\ &&
\exp \bigl\{-  \frac{N}{2} \Str R^2   -N \Str \log (E-R +i \epsilon )   \bigr\} \ .
\end{eqnarray} 

This $R$ integral as $N \to \infty$ should be accessible to saddle point analysis, since
the integrand is of the form $e^{- N S(R)}$ with $N$ large in front of the action $S(R) =
 \frac{1}{2} \Str R^2   + \Str \log (E-R +i \epsilon )   $. The saddle points satisfy
in the limit $\epsilon \to 0$ 
\begin{equation}
R = (E-R)^{-1} \  ;  \ R^2 - ER +1 =0
\end{equation} 
so the corresponding eigenvalues are
\begin{equation}
\cE = E/2 - i \sqrt{1-E^2/4} \ , \   \cE^* = E/2 + i \sqrt{1-E^2/4},  
\end{equation} 
with $\cE \cE^* = 1$, $E-\cE = \cE^*$. The corresponding bosonic saddle points are
\begin{equation}
 R = E/2 - i \sqrt{1-E^2/4}  \left(  \ba{cc}   s_a & 0 \\  0 & s_b \ea \right).
\end{equation} 
with $s_a= \pm 1, s_b = \pm 1$
These saddle points are not along the original path of integration, which is $a$ real and
$b$ real. So we have to shift the $a$ integration contour to include
an imaginary part $-is_a \sqrt{1-E^2/4}  $ and the $b$ contour to include
an imaginary part $-i E/2$. We remark that in order to avoid crossing singularities
we must take into account the positive sign of $\epsilon$
and choose $s_a = +1$ if $\epsilon >0$.  So we are left with two saddle points
at $s_b = \pm 1$. But it is possible to explicitly perform the Fermionic integration
and check that the saddle point at $s_b = 1$ dominates the saddle point at $s_b = -1$,
which is suppressed by a $1/N$ factor. 

Indeed we have
\begin{equation}
(E-R)^{-1}_{bb} =  \frac{1}{E-a}  [1 - \rho^* \rho \frac{1}{(E-a)(E-ib)}]^{-1}\ ,
\end{equation}  
\begin{equation}
\Sdet (E-R) =   \frac{E-a}{E-ib} [1 - \rho^* \rho \frac{1}{(E-a)(E-ib)}] \ .
\end{equation}  
Hence 
\begin{eqnarray}
< \nu (E) > &=& - \frac{1}{\pi }  \Im  \int da db d\rho^* d\rho  
e^{-N/2 (a^2 + b^2 + 2 \rho^* \rho)} \nonumber\\ &&
 \frac{(E-ib)^N}{(E-a)^{N+1}}  [1 - \rho^* \rho \frac{1}{(E-a)(E-ib)}]^{-N-1} \ ,
\end{eqnarray} 
so that, performing exactly the $\rho$ and $\rho^*$ integration, and translating the contours, we get,
remembering that by (\ref{grass}) and anticommutation $\int d\rho^* d\rho  \rho^* \rho = -\frac{1}{2\pi} $:
\begin{eqnarray}
< \nu (E) > &=&  \frac{1}{2\pi ^2 }  \Im  \int_{\Im a = -\sqrt{1-E^2/4}} da 
\int_{\Im b =-E/2} db  e^{-N/2 (a^2 + b^2)}  \nonumber\\ &&
 \frac{(E-ib)^N}{(E-a)^{N+1}}  N [ -1 +  \frac{1}{(E-a)(E-ib)}  + O(1/N)] \ .
\end{eqnarray}
At the vicinity of the first saddle point we have $a=\cE + \delta a$, $ib = \cE + i \delta b$
and by ordinary Hessian analysis we need to expand to second order in $\de a$ and
$\de b$. We find
\begin{equation}
 e^{-N/2 (a^2 + b^2)}  \frac{(E-ib)^N}{(E-a)^{N}} =   e^{-N/2 (1-\cE^2)(\delta a^2 + \delta b^2) + O(\delta^3)} \ .
\end{equation} 
Furthermore at the first saddle point we have $[ -1 +  \frac{1}{(E-a)(E-ib)} ]= -1 + \cE^2 $
and $ \frac{1}{(E-a)} = \cE$
so that the Hessian approximation near the first saddle point gives a contribution
\begin{eqnarray}
< \nu_1 (E) > &=&  \frac{1}{2\pi ^2 }  \Im \int  \delta a 
\int  \delta b   e^{-N/2 (1-\cE^2)(\delta a^2 + \delta b^2)} \cE
(-1+ \cE^2   + O(1/N))  \nonumber\\
&=& \frac{1}{\pi } \sqrt{1-E^2/4} + O(1/N)\ .
\end{eqnarray} 
At the second saddle point we have $a=\cE + \delta a$, $ib = \cE^* + i \delta b$
so that $-1 +  \frac{1}{(E-a)(E-ib)} = 0$ at this other saddle point. Therefore to leading
order as $N\to \infty$ there is no contribution to $\nu(E)$ from this other saddle point.
It is possible to bound the contributions away from the saddles and to conclude therefore
\begin{equation}
<\nu(E)> =  \begin{cases}\frac{1}{\pi } \sqrt{1-E^2/4}   \quad {\rm for}\  |E| \le 2 \ ,\\
0 \quad {\rm otherwise}\ . \end{cases}
\end{equation} 

The computation of level correlations is similar but complicated by
lack of positivity of the Hubbard-Stratonovich form, which leads
to technical complications. There is also an additional symmetry,
and the main non-trivial mean field integral therefore takes values
in a coset superspace parameterized by eight real variables, four bosonic and four
Fermionic \cite{Mir}.

Let us also stress that the corrections as $N \to \infty$
to Wigner's law can be in principle systematically computed
but the computation becomes more and more difficult in practice
for higher order terms. However it is relatively easy to derive e.g.
crude bounds on the probability that a random $N \times N$
matrix develops a large norm using some kind of Tchebycheff inequalities.
For instance it is an instructive exercise to prove \cite{Poir}

\begin{lemma}
For sufficiently large $a$ and $N$ one has
\begin{equation}
{\rm Prob}  (\Vert H \Vert  \ge a \sqrt{6})  \le 8Na^6/27  e^{-N^{1/3} a^2 /3} \ .
\end{equation} 
\end{lemma}
The Tchebycheff inequality here is simply to use
\begin{equation}
{\rm Prob}  (\Vert H \Vert  \ge a \sqrt{6})    \le (6a^2)^{-m} <  \Tr H ^{2m} >
\end{equation} 
and to optimize over $m$. Gaussian integration for $<  \Tr H ^{2m} >$ can
be explicitly performed through Wick's theorem. It leads to the evaluation of
so called Feynman graphs. A careful counting of these graphs leads to the proof
of the lemma by taking $m$ proportional to $a^2$. This bound is not optimal but shows clearly
that the probability for a random matrix normalized in this way to develop eigenvalues much larger
than $O(1)$ is very small. Such crude bounds are sufficient for complete control of
the Anderson model in the regime $|p^2-E | >> \la^2$, see below.

\section{The Anderson Model}

The Anderson model of an
electron in a random potential corresponds to the Hamiltonian
$$
H = -\Delta + \la V(x),
$$
acting on the Hilbert space $L^2 (\RR^2)$, where $\Delta$ is
the usual Laplacian and $V$ is a real Gaussian process on
$\RR^2$ with short range correlations. $\la$
is a coupling constant that allows to adjust the strength of
the random disorder with respect to the deterministic part. 

This model was initially proposed to describe the electron 
motion in doped semi-conductors at low temperature or in 
normal disordered metals. It is now the central model for 
the theory of electronic transport and wave propagation in 
disordered systems~\cite{EFSHK}. It was conjectured by 
Anderson as soon as 1958~\cite{And58} that such a model 
exhibits a localized phase in which the electrons are 
trapped by the defects. In 1979 it was argued that this 
model has a phase transition in dimensions three or more 
between the localized phase and an extended one~
\cite{gang4}.

The localized phase is now well under control. In one 
dimension, localization was rigorously established for any 
disorder at the end of the seventies~\cite{GMP,KuSo}. Later 
localization was established in any dimension at strong 
disorder or for energies out of the conduction bands
\cite{FMSS}. A simplified and more efficient method to get 
this result was given in~\cite{AiMo}.

In contrast the weak disorder regime is still poorly
understood. In two dimensions it has been argued~%
\cite{gang4,Bergmann} and numerically established~\cite{Kra} 
that localization persists at arbitrarily small disorder,
with a localization length diverging like 
$O(e^{c/\la^{2}})$. In dimension three, numerical 
simulations confirm the existence of the Anderson
transition~\cite{Kra}, leading to an extended phase. In 
addition, analytical results~\cite{Efet83} and other 
numerical calculations show that the level spacing 
distribution follows the Wigner-Dyson distribution for 
random matrix theory (RMT)~\cite{Kra2}. This gave the
motivation for a description of mesoscopic system in terms 
of RMT~\cite{AltShk}. This method has been very successful 
when compared to experiments and was the source of 
developments of supersymmetric methods~\cite{Efet,Mir} in 
solid state physics. But this heuristic connection between random matrices
and the Anderson model remained mysterious.
And on the rigorous level it is still a 
mathematical challenge up to now to prove even regularity of 
the DOS at weak disorder in the conduction band (that is, \ analyticity in energy in a narrow band around 
a real interval in the conduction band of the deterministic Hamiltonian, see Conjecture 1 below).
This is still unknown in either dimensions
two or three (see \cite{CHKR} for recent continuity results outside the conduction band). 

Using a phase-space analysis inspired by the renormalization
group method around Fermi surfaces in condensed 
matter~\cite{BG,FT,FMRT}, G. Poirot and coauthors~ \cite{MPR1}
have understood the connection between the Anderson model
and random matrices. They established that the effective Hamiltonian near the
Fermi level is given indeed by a random matrix model. In the simplest case, that of
two dimensions this random matrix model is almost identical to the GUE
(Gaussian Unitary Ensemble), but just contains one extra discrete
symmetry. This symmetry is called the Flip Symmetry
in \cite{BMR}, and the corresponding Flip Matrix model
has been analyzed through the supersymmetric method.
In three dimensions 
the flip symmetry becomes a continuous $U(1)$ symmetry, and produces more
complicated correlations between matrix elements~%
\cite{MPR1}; however even in that case, these correlations are not expected
to change the main statistical properties of the spectrum, such as Wigner's law
for the density of states.

In the coming subsections we will summarize the content of \cite{BMR}.
Let us recall that in \cite{BMR} the flip symmetry was slightly simplified. This
flip symmetry is indeed essentially a $\mathbb Z_2$
symmetry, but which effectively degenerates into a larger symmetry near the diagonals of the matrix,
which correspond to degenerate rhombuses in the momentum representation.
Therefore in the last section of this review we will refine the phase space analysis of the Anderson model
and propose improved matrix models
to study this slight degeneracy near the diagonals of the 
flip symmetry, using some hierarchical approximations that mimic
the true situation but are more tractable from
the point of view of the supersymmetric mean field analysis. Similar ideas could also apply to the study of the  $U(1)$ three dimensional symmetry.

\subsection{Phase Space Analysis}

As in other condensed matter models, the ultraviolet
region in momentum space is irrelevant and cut off in the Anderson model. 
The most common way to do this is to replace ordinary space by a lattice such
as $\mathbb Z ^d$. This corresponds physically to the so-called
tight binding approximation in which the electron spends all
its time on  lattice sites in a crystal, jumping from one site to an other. 
This lattice model automatically cuts off large momenta since the Fourier
transform of the lattice $\mathbb Z ^d$ is simply the torus
$[-\pi , \pi]^d$.

Since the lattice cutoff breaks rotation invariance,
it is also common to consider a rotation
invariant momentum cut off. For that we consider the free Hamiltonian without
the random potential, which is simply the Laplacian in a suitable unit system:
\be   H_0 = - \Delta 
\ee
and the free retarded Green's function:
\be   G_0 = \frac{1}{H_0 -E -  i \epsilon}.
\ee
also called the {\it propagator} in physics.
In Fourier space this is a diagonal multiplication operator. A rotation invariant
momentum cutoff can be implemented by multiplying this operator by a smooth
$C^{\infty}$ function $\kappa$ with compact support which suppresses
large momenta. For instance we can ask that this function is identically 1 for $|p| \le A$
and 0 for $|p| \ge A+1$, where $A$ is a fixed number, e.g. 10.
This leads to  define the cutoff free retarded Green's function:
\be   G_{0,\kappa} (p)= \frac{\kappa(p)}{p^2 -E -  i \epsilon}  .
\ee
This is usually called in physics the "jellium" cutoff, and we call the corresponding
model the jellium Anderson model.
Regular perturbation theory for this model consists in computing the 
interacting Green's function as a power series in $\lambda$ through
a resolvent expansion:
\be G = G_0 \frac{1}{1+\la V G_0} = G_0 -G_0\la V G_0 + G_0 \la V G_0 \la V G_0 + ...
\ee
The rotation invariant cutoff consists in replacing everywhere
$G_0$ by $G_{0,\kappa}$. Remark that the convergence of the power series
for fixed $V$ is no problem at $\la$ small enough, since this is a geometric series.
However we are not interested in any particular $V$ but on the average over $V$.
The corresponding integrated Green's function is obtained by integrating
over $V$, using the Gaussian rules of integrations, called Wick's theorem
in physics.  Only terms with even numbers of $V$ survive and
the result can be indexed by drawings called Feynman graphs. In these
Feynman graphs, the $G_0$ propagators can be pictured as full lines,
whether the covariances for contracted $V$, which are $\delta$ functions in $x$-space,
are usually represented by dotted lines (see Figure 1).
\begin{figure}
\centerline{\psfig{figure=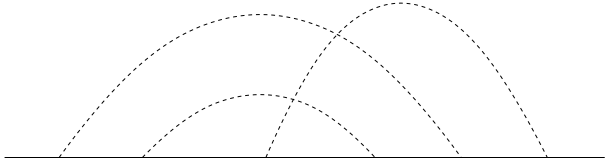,width=8cm}}
\caption{A typical Feynman graph}
\end{figure}
In this case the first non trivial graph also called tadpole graph 
(see Figure 2)
\begin{figure}
\medskip
\centerline{\psfig{figure=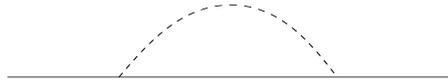,width=6cm,height=1cm}}
\caption{The first non trivial graph, the "tadpole"}
\end{figure}
gives a momentum space contribution
\be   C = \la^2 \int d^{d}p  \frac{\kappa(p)}{p^2 -E - i \epsilon}  .
\ee
This is a pure number, with a finite imaginary part $\de$ as $\epsilon \to 0$. For instance
in two dimensions and for $E=1$, this imaginary part $\de$ is $\la^2\pi^2$. 
Therefore physicists, resumming only the
graphs $G_n$ with $n$ consecutive tadpoles
(see Figure 3), which form a geometric series,
\begin{figure}
\centerline{\psfig{figure=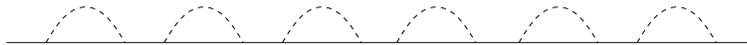,width=10cm}}
\caption{The graph $G_n$ with n tadpoles}
\end{figure}
and neglecting all other graphs,
can perform the limit $\epsilon \to 0$:
\be  G_{tadpole} = G_0\sum_{n=0}^{\infty}   (\de G_0)^n
= \frac{\kappa(p)}{p^2 -E -  i \de\kappa(p)}  . \label{tadapp}
\ee
With their usual boldness, they conclude that the averaged Green's function should be regular in momentum space, and consequently also decay in $x$-space at a spatial 
scale inverse of $\de$, at least for small $\la$.
The corresponding conjectures, alas unproved yet, could be described as follows:

\medskip
\noindent{\bf Conjecture 1: Analyticity of the Averaged Density of States}

{\it 
Let $[a,b]$ is a fixed interval well inside the spectrum of the unperturbed Laplacian spectrum (this
means $[a,b]\subset ]0,2d[$ for the lattice model, and $[a,b]\subset  ]0,A[$ for the jellium model).
For any $\la$ small, the averaged density of states $<\nu (E)>$ is analytic in $E$ in a rectangle
centered on the real interval $[a,b]$ and of width $O(\la^2)$, 
hence of type $[a,b]\times[-i O(\la^2) , +i O(\la^2)]$.
}

\medskip
\noindent{\bf Conjecture 2: Scaled Polynomial Decay of the Averaged Green's Function}

{\it There exists a constant $c$ such that for any integer $q$, there exists a constant $K_q$ such that the averaged Green's function $<G>$ in direct space obeys long range spatial decay of 
power $q$ and scale $O(\la^2)$:
\be  \vert <G(x,y)> \vert  \   \le\  K_q (1+ c \la^2 |x-y|)^{-q}\ .
\ee
}
\medskip

These conjectures are not completely optimal, but even in this form they are still a challenge.
It is in fact expected that we have analyticity in a rectangle with $[a,b]$ 
closer and closer to the full unperturbed spectrum as $\la \to 0$.
Also in the case of the lattice model one expects in fact exponential
scaled decay of order $\la^2$. Since a compact support cutoff $\kappa$ cannot be analytic, 
its Fourier transform does not decay exponentially. So for the jellium model one expects
at most fractional exponential decay (if the function $\kappa$ is Gevrey). 

Perhaps more interesting than the conjectures themselves
is what they mean: namely that no matter how small the interaction,
it regularizes the average Green's function. This is somewhat similar
the phenomenon of dynamic mass generation in field theories like QCD, although
here the regulator is polynomial in $\la$. 

These conjectures would be important also
to justify perturbation theory and the tadpole
approximation in the small coupling regime of the Anderson
model. As is well known the ordinary perturbative approach
is plagued by mathematical problems. Indeed due to the large number of
Feynman graphs (there are $n!!= (n-1)(n-3)...5.3.1$ graphs at order $n$) the integrated
perturbation series has zero radius of convergence.
Over the years, techniques
have been developed to overcome this problem by replacing the complete divergent Feynman series
by carefully truncated expansions with Taylor remainders that can be rigorously bounded.
This set of techniques goes under the name of "constructive field theory" \cite{constr}.

Constructive field theory tells us that in order to achieve non-perturbative
theorems in this kind of situations where the interplay between propagation and
interaction is non-trivial, the best frame is to analyze the theory in phase-space.
Some multi-scale {\it renormalization group} analysis 
 is usually necessary. Renormalization group is the tool
which computes long range behavior from local physical interactions. 
Cluster and Mayer expansions are the correct non-perturbative steps
which make the renormalization group mathematically well-defined \cite{constr}. 
The guiding line is to identify the relevant degrees of freedom of the problem in phase space,
and then to perform some carefully controlled perturbation theory which roughly derives at most
one perturbation step for each such degree of freedom.

A Hamiltonian, whether random or not, acts indeed on a Hilbert space
of quantum states. Using a system of units in which $\hbar =1$ a quantum 
{\it degree of freedom} corresponds roughly to a unit volume
in phase space. Although the quantum Hilbert space for a particular problem
may be formally infinite dimensional (such as $L^2(\mathbb{R})$...), 
in practice any physical problem should in fact
be well approximated by a finite dimensional version of this Hilbert space.
The main problem is to identify 
the relevant degrees of freedom of the system, and to control their
proliferation in the ultraviolet (field theory) or infrared (statistical mechanics) limit.  

A partition of the classical phase space into cells of
unit volume may be viewed as giving a particular basis 
of this finite dimensional
Hilbert space. Not all partitions are admissible;
they should roughly respect the Heisenberg uncertainty principle
ultimately related to the symplectic geometry of phase space.
Any product of the typical dimensions of any cell in the 
directions of a conjugate pair of positions and momenta should not
be much smaller than 1.

Some decompositions may be more convenient than others for
particular problems. In the lattice Anderson model on a lattice, the phase space
is $\mathbb{Z}^d \times [0,2\pi]^d$ since positions lie on a square lattice and
momenta lie in the dual tori. In the jellium Anderson model,
it is $\mathbb{R}^d \times B(0,A+1)$, where the closed ball $B(0,A+1)$ in $\mathbb{R}^d $
with center 0 and radius $A+1$
is the compact support of the cutoff function $\kappa$. This is not very different 
from the previous case since
$\kappa$ restricts the large momenta, and the $x$-space should be again well
approximated by a lattice of cubes of side size roughly the inverse of $A$, hence of
order $O(1)$. Indeed all propagators and functions built out of them will be roughly
constant within each of these cubes.

The initial natural partitioning of such a phase space is to associate
to each lattice site or cube in direct $x$-space a {\it degree of freedom}. 
Momentum space is compact, and the only reason for which the problem
has infinite number of degrees of freedom is the infrared problem, 
or thermodynamic limit: we want to understand the theory in infinite
volume in $x$-space.

This obvious partitioning is well adapted to the study of the
localized phase of the Anderson model, in which 
the random potential dominates over the propagation.
Indeed a random identically distributed
potential $V$ is simply a multiplication operator in $x$-space. It is
a diagonal operator in $x$ space with i.i.d. diagonal coefficients
$V(x)$ for $x \in \mathbb{Z}^d$. Spectrum is pure point and the eigenvalues of this operator
are obviously the wave functions concentrated on single sites, or $\de$
functions in $x$ space. The
density of states and statistics of eigenvalues are Poissonian: the
presence of an eigenvalue at some energy $E$ simply means
that for some $x$ we have $V(x) = E$, and this information has no influence on
the values of $V(y)$ for $y \ne x$. Therefore in this case
$R_2(s) \simeq_{s \simeq 0} \delta (s) +1$.

Most of the rigorous work on the Anderson model is concerned with the localized
phase. In this phase the addition of a hopping term such as a lattice Laplacian is a 
small perturbation which does
not fundamentally modify the spectrum, which remains pure point,
and the localized character of the eigenfunctions.

But this partitioning of phase space is not the right one to study the other
regime in which it is the lattice Laplacian with its continuous spectrum 
which is the main effect, and the random potential
which should be the perturbation. 

From now on let us restrict to a single averaged Green's function or the density
of states, leaving the level correlations for future studies.  To prove
analyticity in the rectangle of Conjecture 1, we need only to consider
a fixed value $E_0$ in $[a,b]$ and, translating $E$ to $  E_0 + E $, to prove
analyticity for $|E|\le O(\la^2)$. If we can do this for any $E_0 \in [a,b]$, Conjecture 1 follows.
Since this will not change anything  except a global rescaling we always choose $E_0 =1$
in what follows.

If we take seriously the propagator as a guide to the right
partitioning of phase space, we recognize that the interesting region 
in momentum space is the singular region $p^2=1$. This is the d-dimensional
sphere, for instance in two dimensions it is the circle of radius $1$.
We know from the theory of renormalization group around extended singularities \cite{FT,FMRT}
that in this case we should perform a change of basis, dividing the region 
$p^2\simeq 1$ into smaller and smaller cells to analyze it with great care; correspondingly
$x$ space should be cut into dual boxes that will be larger and larger as we approach the singularity,
hence correspond to longer and longer distance effects.

This decomposition correspond to cut first momentum space into shells according
to a geometric progression which pinches more an more the singularity.
Fixing a rate $M>1$ for this geometric progression (for instance $M=2$),
this means that we write
\be \kappa (p) = \sum_{j=0}^{j_{max}} \kappa_j (p)   \label{slices}
\ee
where $\kappa_j (p)$ is a smooth function with compact support which roughly
ensures $|p^2 -1|\simeq M^{-j}$.

The index $j$ plays the role of a renormalization group index, and the corresponding 
region in momentum space is called the $j$-th slice or momentum shell.
The last index $j_{max}$ is chosen so that $M^{-j_{max}} \simeq  \de = \pi^2 \la^2$.
Indeed the expansion should not be infinite. The last function $\kappa_{j_{max}} (p)$ 
for the last slice $j_{max}$ should be different, meaning only $|p^2 -1|\le M^{-j_{max}}$.
The reason is that we expect the basic physical intuition behind (\ref{tadapp}) to be correct.
Averaged Green's functions should decay after that scale; equivalently the singularity
of the free propagator should be screened by the appearance of an imaginary part
due to the interaction with the random potential.  

This screening phenomenon, like mass generation 
under influence of interaction in field theory,
can only occur at a scale where interaction effects
become of the same size than the free propagator. The conclusion is that 
we must distinguish two regimes: 

- The regime $|p^2 - 1 | >> \la^2$, or $j \le j_{max}$. In this region 
the free propagator should dominate over the interaction and regular cluster
and Mayer expansion should prove that this part of the Green's function remains
in every respect close to the free one.

- The regime $|p^2 - 1 | >> \la^2$ or $j \simeq j_{max}$, which corresponds to a 
few renormalization group slices (or in fact just one if we take $M$ big enough).
In this last slice clearly ordinary perturbation theory (even recast under the form of
cluster or Mayer expansion) cannot work since propagators and typical random potentials
are of the same size (otherwise the non-trivial effect of screening would be impossible). 

In this regime the functional integral oscillates wildly.
A first attempt to exploit the corresponding cancellations
through Ward identities was made in \cite{MPR2}.
However this turned out to be a difficult and not completely conclusive approach.
Oscillating integrals with analyticity properties can often be better analyzed by
finding saddle points and shifting contours to pass through them. 
In favorable cases the leading part of the integral then comes 
simply from a Hessian approximation near a single dominant saddle.
This saddle becomes a new expansion point for the theory, similar to a change
of vacuum in field theory. The corresponding shift is definitely a non-perturbative tool,
very difficult to understand in terms of combinations of the initial Feynman graphs 
of Figure 1.

Such a shift precisely occurs in the supersymmetric
formalism of the previous section: the
contour for the mean fields $a$ and $b$ has been shifted to pass through 
a non trivial saddle point that lied out
of the original integration contour. In this way a highly oscillating integral,
has been replaced by an equal one which is much simpler to analyze.
We must explain now why the Anderson model is analogous to a matrix model,
so that the same strategy should also work there.

\subsection{Sectors and the Random Matrix Approximation}

Momentum slices or shells without further decomposition
are not suited for a phase space analysis. Indeed
a free propagator restricted to such a shell has no simple dual decay properties
in direct space. The correct phase space analysis must split further each shell into angular
sectors, along the directions tangential to the extended singularity. What limits the length
of the sectors in these directions is the curvature of the singularity \cite{FMRT}; (see also \cite{R}
for a non-spherical example).
In the case of a spherical singularity such as the one of the jellium Anderson model, 
there are two natural and acceptable solutions to the splitting into sectors:

- split the spherical shell of width $M^{-j}$ into roughly $M^{(d-1) j}$ sectors
(noted by Greek letters such as $\alpha$ or $\sigma$) with same size in all directions. This is the most natural idea
and the corresponding sectors are called isotropic. It is however not optimal
for momentum conservation analysis. The direct space is
then cut into a lattice ${\cal D}_j$ of cubic boxes $\Delta$ of side size $M^{j}$ 
for each value of $j$. 

-  split the spherical shell of width $M^{-j}$ into roughly $M^{(d-1) j/2}$ sectors,
still noted $\sigma$, of length $M^{-j/2}$ in all directions tangent to the singularity. These
sectors are called anisotropic. This is the optimal tool to analyze e.g. momentum conservation,
but the price to pay is that the dual decomposition of $x$-space gives rise to
anisotropic parallelepipeds which change as the sector changes. For each
sector $\sigma$ in a shell $j$ we have a different lattice ${\cal D}_{\sigma}$ of spatial
boxes covering $\mathbb{R}^d$, which have length $M^{j}$ in the direction of
$\sigma$, and  length only $M^{j/2}$ in the other orthogonal directions (since for the sphere
these orthogonal directions correspond to the tangent directions at $\sigma$).

In the simple isotropic case, a phase space cell of roughly unit volume is therefore 
made of a value of $j$
in $[0, j_{max}(\la)]$, a spatial box $\De \in {\cal D}_j$ and a momentum cell 
$\sigma$ in the $j$-th shell. The true Hilbert space of the theory is well approximated by a 
finite dimensional Hilbert space whose basis is indexed by these phase space cells.

Now at last comes the analogy with random matrices.
In $x$ space, $V$ is a real diagonal operator; therefore in $p$ space 
$V(p)$ is complex with 
\be V(-p) = \bar V(p), \label{hermitian}
\ee 
and $V$ is a random convolution operator $Vf (p) = \int dq V(q) f (p+q)$ with 
covariance $<V(p) \bar V(q) > = \delta (p-q)$. A delta or short range
covariance in $x$ space translates into a white noise in $p$ space,
that is all momenta correspond to independent variables with a priori equal weight. 
If we use isotropic sectors to simplify,
it is natural to discretize the momentum space for $V$ with 
cells of the same dimensions that the angular sectors used
for the propagator. When two propagators with momenta
$\alpha$ and $\beta$ join at a potential $V$, the momentum of $V$, or "transfer momentum" by overall 
translation invariance of our theory, is $p= \alpha - \beta$.  There is therefore an effective
ultraviolet cutoff on $V$ (because $\alpha$ and $\beta$ being bounded, so is $\alpha - \beta$).

Here is a first subtlety. There is no reason for $\alpha$ and $\beta$ to lie in the same
shell. The operator $V$ can be decomposed into the shell-conserving
or diagonal part $V_{s}= \sum_{j} V_{s,j}$ and the shell-changing or off-diagonal part $V_{o}$.
For the time being, let us neglect $V_o$ and consider an operator $V_{s,j}$ acting
in a single shell.

The covariance for $V$ has no singularity and we should therefore discretize
the momenta for $V$ into roughly $M^{dj}$ cells $\tau$ of side size $M^{-j}$. 
Hence if we divide space according to the lattice ${\cal D}_j$, to a given
cube $\Delta$ should correspond roughly $M^{dj}$ discrete variables
$V_{\tau, \Delta}$, which are roughly independent and identically distributed.
The interaction being local in $x$ space, such a variable  $V_{\tau, \Delta}$
must join sectors $(\alpha, \Delta)$ and $(\beta, \Delta)$
spatially localized in the same cube $\Delta$ of ${\cal D}_j$. From now on fixing such
a cube $\Delta$, we can therefore consider the variables $V_{\tau, \Delta}$,
as the discrete elements of a random matrix that we call $V(\Delta)$. 

Let $\alpha$ and $\beta$ be two indices parametrising two sectors of the same shell $j$.
We need to know which pairs of sectors a given independent variable $V_{\tau, \Delta}$ can join,
so to analyze the relation $\tau = \alpha - \beta$.
The answer happens to be strongly dimension-dependent.

A first remark is that if $\tau = \alpha - \beta$, then $-\tau = \beta - \alpha$.
This means, using the relation (\ref{hermitian}) that the matrix $V$ must be hermitian,
in any dimension:
\be  V_{\alpha, \beta} = \bar V _{\beta, \alpha}\ . \label{fl1}
\ee
In two dimensions if we consider a given $\tau$ of length between 0 and 2$E$,
there are typically roughly two pairs of momenta of length $E$, forming a rhombus
such that $\tau = \alpha - \beta$. These two pairs are $(\alpha, \beta)$
and $(-\beta, - \alpha)$. Therefore the matrix $V$ in addition to being hermitian
has an additional symmetry
\be   V_{\alpha, \beta} = V_{-\beta, -\alpha} \label{fl2}
\ee
called the flip symmetry in \cite{BMR}. Under this symmetry, two coefficients
in e.g. the upper triangle of the matrix correspond to the same random variable.
This means that $V$ is not exactly a GUE. We can label the sectors of the circle
in such a way that these two identified coefficients are symmetric with respect
to the antidiagonal of the matrix. For that it suffices to divide the circle into an even
number $2N$ of sectors labeled as 
$\{1,\cdots,N\} \cup \{-N, \cdots,  -1\}$, so that 
$\{1,\cdots,N\}$ labels projective sectors, as shown in
Figure 4.
\begin{figure}
\centerline{\psfig{figure=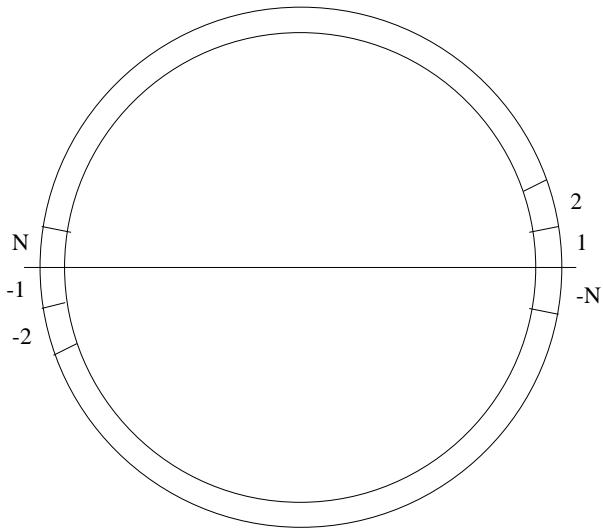,width=8cm}}
\caption{Numbering sectors}
\medskip
\end{figure}
Of course this is only a generic situation, and when the rhombus degenerates
the answers are different. For instance for a momentum $\tau$ of length 2,
there is only one pair $(\alpha, \beta) = (\tau/2, - \tau/2)$ such that 
$\tau = \alpha - \beta$, and for a momentum $\tau$ of length 0, every pair
$(\alpha, \alpha)$ is a solution. This means that all coefficients on the diagonal
of the matrix $V$ are equal. Intermediate situations when the rhombus almost degenerates
should also complicate the flip symmetry, and these issues are discussed in some detail
below.

In three dimensions for a given $\tau$ there is typically an orbit of $\sqrt N$ pairs of
sectors $(\alpha, \beta)$ such that $\tau = \alpha - \beta$, forming a cone. The discrete
$Z_2$ flip symmetry is therefore replaced by a $U(1)$ symmetry; the set of upper 
diagonal coefficients of the matrix $V$ are therefore split into  $O(N^{3/2})$ orbits under this
symmetry, each orbit containing roughly $\sqrt N$ coefficients. Clearly this matrix model
is farther from the GUE model, but still the density of states should fall in 
the same category.

Having recognized the analogy, in phase space, between the Anderson model and 
random matrices, we shall now review first the regime $|p^2 - E | >> \la^2\pi^2$, then
the last shell $|p^2 - 1 | \simeq \la^2\pi^2$.

\subsection{The regime $|p^2 - 1 | >> \la^2\pi^2$}

This is the region in phase space where by definition the deterministic 
part $-\Delta - 1$ is forced to be not too small, and where the random part $\la V$ 
is statistically very small compared to this deterministic part. 
However the distribution for $V$ is not compactly supported. Therefore the probability
for $V$ to have a norm comparable in size to $- \Delta -1$ is never zero,
even in that regime. However it is so small that if such an event 
happens in a phase space region $R$, there is such a small associated
probability factor that it can pay for a very crude bound, obtained by translating
directly the integration contour for the potential $V$ in that region $R$. In this way we
can prove not only that such an event has small probability, but that it does
not influence other regions of phase space.

The corresponding rigorous analysis has been performed in great detail in
the work of G. Poirot \cite{Po}, and here we give only a rough sketch of the argument.

To get an idea of the respective sizes of the deterministic and random pieces
in $( H - 1 )$ we can first work at fixed $j$ for a shell-conserving piece $V_{s,j}$. 
Consider first the deterministic part $p^2 - 1$. Restricted to the $j$-th shell it is 
by definition of that shell an operator with a norm of order $M^{-j}$.
We want to prove that when restricted to that shell, the random part has typically
a much smaller norm. If this is the case ordinary perturbation theory of the random
part should be trusted.

The full interacting averaged Green's function is
\be  \frac{\kappa(p)}{p^2 - 1-E+i\epsilon }(1+ \la V \frac{\kappa(p)}{p^2 - 1-E+i\epsilon } ) ^{-1} \ .
\ee
We have divided into shells through $\kappa = \sum_j \kappa_j$. For a slightly more
symmetric form we can put square roots of the free propagator on each side of $V$,
and write $\kappa_j =\eta_j^2$. Then 
the random operator restricted to the $j$-th shell is
\be  H_j(p,q) = \frac{\eta_j (p)}{
 (p^2 - 1 -E +i\epsilon ) ^{1/2}} \la V(p-q) \frac{\eta_j (q)}{(q^2 - 1 -E +i\epsilon ) ^{1/2} }
\ee 
with
\be  V(p-q) = \int dx  e^{i x.(p-q)}V(x) \ .
\ee
The main problem to define and control the averaged Green's function
is now to compare the norm of $H_j$ to 1 in order to invert
the operator $1+H_j$.

The sectors in a cube are roughly $L^2$ normalized
characteristic functions $e_{\al} = M^{dj/2}\ch_{\al} (p)$, where the factor $M^{dj/2}$
is required for normalization.

The operator $H_j$ restricted to a spatial cube of size $M^j$
is now analogous to a matrix between sectors
with coefficients
\begin{eqnarray}  H_{\al, \beta} &\simeq &  < e_{\alpha} , H e_{\beta} > \\
&=& M^{dj} \int dp dq
\frac{\ch_{\al} (p)}{ (p^2 - 1-E  +i\epsilon ) ^{1/2}} \la V(p-q) 
\frac{ \ch_{\beta} (q)}{(q^2 - 1-E  +i\epsilon ) ^{1/2}}\nonumber
\end{eqnarray}
since $\eta_j \ch_{\alpha} \simeq \ch_{\alpha}$, $\eta_j \ch_{\beta} \simeq \ch_{\beta}$.
In a shell of index $j$, we have roughly $(p^2 - 1  +i\epsilon ) ^{-1/2} = M^{j/2}$, so that
the covariance between coefficients can be computed as
\begin{eqnarray}
<  H_{\al, \beta} ,  \bar H_{\ga, \delta}  >_{V} &\simeq & 
M^{(2d+2)j} \la^2 \int dp dq dp' dq' \ch_{\al} (p)\ch_{\beta} (q)
\ch_{\gamma} (p')\ch_{\delta} (q')\nonumber\\
&& \int dx dy   e^{i x.(p-q)}e^{-i y.(p'-q')} < V(x) V(y) >_{V}\ .
\end{eqnarray}
Using $< V(x) V(y) >_{V} = \delta(x-y)$ we find
\begin{eqnarray} <  H_{\al, \beta} ,  H_{\ga, \delta}  >_{V} 
&\simeq&  M^{(2d+2)j} \la^2 \int dp dq dp' dq' 
\nonumber\\
&&\ch_{\al} (p)\ch_{\beta} (q)
\ch_{\gamma} (p')\ch_{\delta} (q') \delta((p-q)- (p'-q')) 
\nonumber\\
&\simeq&  M^{(2d+2)j} \la^2 \int dp dq dp' \ch_{\al} (p)\ch_{\beta} (q)
\ch_{\gamma} (p')\ch_{\delta} (p'-p+q) \nonumber\\
&\simeq &  \la^2  (\delta_{\alpha, \gamma} \delta_{\beta, \delta} + 
\delta_{\alpha, -\delta} \delta_{\beta, -\gamma}) \ \ {\rm for\ d=2}\; .
\end{eqnarray}
In the last line we specialize to the generic momentum conservation rules
for $d=2$ (non-degenerate rhombuses)\footnote{For general $d$ we would find 
a factor $ \la^2 M^{(2-d)j} $ times the constraint that there are momenta $p$,
$q$ and $p'$ in the sectors $\alpha$, $\beta$, $\gamma$ such that  $p'-p+q $ is in $\delta$,
and this is a more complicated constraint.}.

The random operator $H_j$ in two dimensions restricted to a single cube, 
is therefore similar to a random matrix of the GUE type
with the additional flip or $U(1)$ symmetry, and with covariance proportional
to $\la^2$. The size of the matrix is $N\sim M^{j}$,
the number of isotropic sectors. We 
know the average norm
of a $N$ by $N$ GUE matrix to be not much more than $O(1)$ for matrices
with covariance $1/N$ on the independent coefficients (see Lemma 1 in section 2).
Rescaling to covariance $\la^2$ is like rescaling $H$ by $\la \sqrt{N}\sim  \la M^{j/2}$.
Hence we conclude that 
statistically the norm of the random part $H_j$ should not be much larger than
$O(1) \la M^{j/2}$.
This means it is statistically much smaller than 1 for $\la M^{j/2} << 1$,
which is nothing but  $j \le j_{max} (\la)$ (if $M$ is large).
We conclude that statistically we can perturb the random part with respect to the deterministic part
in the whole regime $|p^2 - 1 |>> \la^2$, as announced.

Of course this is a very rough argument and we must provide details on what to do
in the infinite volume case, and how to treat the statistically rare events where
the random part is unfortunately much larger than expected.

Since the space volume can be large, one performs a battery
of tests which for each cube $\De$ tell us whether the random part in that cube
$\De$ has an exceptionally large norm or not. This 
battery of tests is called a large/small field analysis in constructive theory.

Roughly speaking one uses then an analyticity argument in the cubes where $V$
has anomalously large norm, which form the so-called large field region. 
The contour of integration
for the $V$'s of this large field regions is translated to create an imaginary 
part of same order than the deterministic part, and a bound is applied to
a resolvent expansion which tests whether the Green's function effectively
or not has visited these large field regions \cite{Po}.

Taking absolute values for this bound destroys oscillations so we need to pay
for a large factor corresponding to the apparent change in the Gaussian normalization:
\be  |e^{-(V+ia)^2/c}| /e^{-V^2/c}  = e^{+a^2/c} \label{cost} 
\ee
but this large factor is compensated by the small probabilistic factor
coming from the large field condition. To understand roughly why this is possible in two dimensions,
where $N \simeq M^{j}$,
recall that to create with $\la V$ an imaginary part of size $M^{-j}$ requires a translation
of the zero momentum $V_{0}$ coefficient on the diagonal
of the matrix $V$ by $i\la^{-1}M^{-j}$. Then the factor
to pay following (\ref{cost})
is $e^{a^2/c}$ with $a=\la^{-1}M^{-j}$ and the covariance $c$ for the
zero momentum coefficient is $M^{-2j}$. So the factor
to pay is $e^{\la^{-2}}$. But if the norm of $V$ is larger than $\la^{-1} M^{-j}$,
by lemma 1, the small probabilistic factor is of order $e^{- M^{j/3} \la^{-2}/3}$
so it compensates for much more than the factor to pay. This is what essentially
fuels the analysis of \cite{Po}.

This large field analysis is also complicated by some auxiliary expansions
to take into account the non-diagonal shell changing parts of the potential,
which does not change the overall picture.
Altogether this analysis, although technically tedious,
is obviously much cruder than a shift of the expansion point, 
which becomes necessary to treat the second regime.

The conclusion is that in this regime one can invert the operator $H-E$,
and the averaged Green's function restricted to that regime has the expected
size and spatial scaled decay rate in $|x M^{-j_{max}}|= \la^2 |x| $. 

\section{The regime $|p^2 -1 |  \le c\la^2$}

We know heuristically from first order perturbation theory 
that the retarded or advanced Green's functions
should decay with a rate of order $\la^2\pi^2$. Therefore it makes sense,
at the level of a single Green's function or for the
density of states to treat this regime as a single momentum slice.
Expecting this decay, it is enough to understand the model
in a single space cube of side size $\la^{-2}$, and then to treat the full model
through some cluster expansion. In this section we want to summarize the 
more detailed computations in \cite{BMR} which prove that the density of
state of a single matrix with the $\mathbb{Z}_2$
flip symmetry has the same large $N$ limit than a GUE matrix. The core of the argument 
reproduces the supersymmetric computation of section 2.3, but with some additional 
auxiliary fields to take into account the flip symmetry. It is then shown that 
the presence of these fields do not modify the main saddle point and the leading 
contribution to the density of states.

\subsection{The Flip Matrix model}

The main idea in \cite{BMR}
is to introduce a discrete version of this single cube problem, in which 
phase space and the 2d momentum conservation rules
have been discretized and somewhat simplified, according to the isotropic
discretization and to equations (\ref{fl1}-\ref{fl2}) of subsection 3.2.

Using the labeling of Fig 1, the structure of $V$ can 
be summarized as follows (remember
rows and columns are labeled as
$\{1,\cdots,N,-N, \cdots,  -1\}$ hence {\it not}
as $\{1,\cdots,N,-1, \cdots,  -N\}$):
{\tiny
\begin{align*}
&\mbox{\normalsize $V$}=
\\
 &\left(
   \begin{array}{cc}
    \left[
     \begin{array}{ccccccc}
        V_0  &&\vdots     &&\vdots && \\
       &\ddots     &\vdots &&\vdots && \\
       &\cdots     &V_0 &\cdots &V_{\al,\beta} & \cdots & \\
      &&\vdots & \ddots & \vdots  &&  \\
       &\cdots   &\overline{V_{\al,\beta}} &\cdots& V_0  & \cdots &      \\
      &&\vdots    && \vdots & \ddots  & \\
      && \vdots   && \vdots && V_0
     \end{array}    \right] &
    \left[     \begin{array}{ccccccc}
      &&\vdots && \vdots &&      \\
       & \ddots    & \vdots && \vdots && \\
       & \cdots    &V_{\al,-\beta} & \cdots  & V_{\al,-\al} 
 & \cdots          & \\
      && \vdots    & \ddots & \vdots && \\
       &\cdots  &V_{\beta,-\beta}  &\cdots  &V_{\al,-\beta}
           &\cdots         & \\
      &&\vdots    && \vdots & \ddots & \\
      && \vdots   && \vdots &&
     \end{array}    \right] \\
--------------------& -------------------- \\
    \left[     \begin{array}{ccccccc}
      &&\vdots  &&\vdots && \\
       &\ddots     & \vdots && \vdots && \\
       & \cdots    &\overline{V_{\al,-\beta}} & \cdots
       &\overline{V_{\beta,-\beta}}  & \cdots & \\
      && \vdots    & \ddots & \vdots && \\
       &\cdots     &\overline{V_{\al,-\al}}  &\cdots
&\overline{V_{\al,-\beta}}   &\cdots & \\
      && \vdots   && \vdots & \ddots  & \\
      && \vdots   && \vdots &&
     \end{array}    \right]
  &  \left[  \begin{array}{ccccccc}
        V_0 &&\vdots &&\vdots && \\
       & \ddots    & \vdots  && \vdots && \\
       & \cdots &   V_0 & \cdots & {V_{\al,\beta}}
 & \cdots          &  \\
      && \vdots    & \ddots & \vdots
          &&     \\
       &\cdots &\overline{V_{\al,\beta}}
 &\cdots & V_0   & \cdots & \\
      && \vdots && \vdots & \ddots & \\
      && \vdots   && \vdots && V_0
     \end{array}    \right]
\end{array} \right) 
\end{align*}
}
and the main result of \cite{BMR} is:
\begin{theorem}
In the large $N$-limit, the DOS of this flip matrix model
converges to Wigner's semi-circle distribution. The
corrections to the limit are uniformly bounded as $O(1/N)$
as $N \to \infty$.
\end{theorem}

Let us summarize now the proof of this theorem.

\subsection{The isotropic 2d flip model: Fermionic terms}

Introducing supersymmetric fields 
$$
    \Psi_{\si \al} = (S_{\si \al} , \ch_{\si \al}),
$$
for $\si = \pm 1$ and Greek variables such as $\al$,
$\beta$ now running only within $[1,...,N]$. 
we find the following formula for the density of states
\be
\label{susy.eq-supsersym1}
\nu (E) = \lim_{\Im E\to 0^+}\frac{-1}{\pi}\Im
\int \frac{i}{2N}\sum_{\al, \si} S^{+}_{\al \si} S_{\al\si}e^{\imath L_0} \prod_{\al,\si}\,
 d\Psi_{\si \al}^{+}\, d\Psi_{\si \al}\, d\mu(V),
\ee
where the supersymmetric action decomposes as
$$
    L_0 = (A_0 + B_0 + C_0),
$$
with
\begin{align*}
    A_0
   &{}=\sum_{\al < \beta,\si} \big\{\la V_{\al,\si\beta}
    (\Psi_{\al}^{+} \Psi_{\si \beta} + \Psi_{-\si\beta}^{+}
        \Psi_{- \al}) + h.c. \big\},
\\
    B_0
   &{}=\sum_{\al} \la V_{\al ,- \al}(\Psi_{ \al}^{+}
    \Psi_{- \al} + h.c. ),
\\
    C_0
   &{}=(E + \la V_0 ) \sum_{\al, \si }
    \Psi_{\si \al}^{+} \Psi_{\si \al}\ .
\end{align*}

The Gaussian integration
over the $V$ variables will be performed except for $V_0$,
leading to the following quartic action
$$
    \nu (E) = \lim_{Im E \to 0^+} \frac{1}{\pi}\,
    Im  \int S^{+}_{\al} S_{\al}  e^{-L}
    \prod_{\al , \si } d \Psi_{\si \al}^{+} \,
        d \Psi_{\si \al} \,d\mu_0 (V_0),
$$
\begin{align}
    L  &{}= \la^2 ( A + B + C), \notag
\\
\label{susy.eq-termA}
    A
   &{}=\sum_{\al < \beta , \si} (\Psi^{+}_{\al}
    \Psi_{\si \beta }+\Psi^{+}_{-\si\beta }\Psi_{-\al})
    (\Psi^{+}_{\si \beta }\Psi_{\al}+\Psi^{+}_{-\al}
        \Psi_{-\si \beta }),
\\
    B
   &{}= \sum_{\al}\Psi^{+}_{\al}\Psi_{-\al}
   \Psi^{+}_{-\al }\Psi_{\al },\notag
\\
    C
   &{}=     (- i\la^{-2} E -  i\la^{-1} V_0 )  \sum_{\al, \si }
    \Psi_{\si \al}^{+} \Psi_{\si \al}. \notag
\end{align}
Some tedious rewriting using the commutation and anticommutation rules
eventually leads to $ L = \la^2 ( \cA + \cB + \cC )$ with
\begin{align}
    \cA
   &{}=(1/2)\sum_{\al , \beta , \si}  ( \Psi^{+}_{\al}
    \Psi_{\si \beta}+\Psi^{+}_{- \si\beta }\Psi_{-\al })
    (\Psi^{+}_{\si \beta }\Psi_{\al } +
        \Psi^{+}_{ -\al }\Psi_{-\si \beta }), \notag
\\
    \cB
   &{}= \sum_{\al} \cB_{\al}, \notag
\\
    \cB_{\al}
   &{}= - 2 S^+_{\al }S_{-\al}S^+_{-\al }S_{\al}
    -\frac{1}{2}\sum_{\si}(\Psi^{+}_{\si\al}
        \Psi_{\si\al})^2 \notag
\\
   &{}\phantom{={}}{}-\Big(\sum_{\si}S_{\si\al}^{+}
    \ch_{-\si\al}^{+}\Big) \Big(\sum_{\si}S_{\si\al}
        \ch_{-\si\al}\Big), \notag
\\
\label{susy.eq-calcdef}
    \cC
   &{}= C = (-i \la^{-2} E-i \la^{-1} V_0) \sum_{\al, \si }
    \Psi_{\si \al}^{+} \Psi_{\si \al}.
\end{align}
The term $\cB$ being a sum of {\it diagonal quartic\/}
terms, will be neglected for a while. For indeed it cannot
be written as the {\it square\/} of a sum over $\al$.
However, the bosonic part of this term has the
{\it wrong sign\/}, which may create some difficulties when
performing the integration in~(\ref{susy.eq-supsersym1}).
This problem will be addressed later.

The other terms can be reorganized so as to get the square
of a sum over the $\alpha$'s by pushing the terms with index
$\alpha$ on the left and the ones with index $\beta$ on the
right. This must be done with care according to the commutation rules for
Bosons and for fermions. The calculation is tedious but
straightforward and gives:
\begin{align}
    \cA
   &{}=(1/2)\Big(\sum_{\al, \si} S^{+}_{\si\al}
    S_{\si\al}\Big)^2-(1/2)\Big(\sum_{\al,\si}
        \ch^{+}_{\si\al}\ch_{\si\al}\Big)(h.c.)\notag
\\
   &\phantom{={}}{}+2\Big(\sum_{\al}S^{+}_{\al}
    S^+_{-\al}\Big)(h.c.)+\Big(\sum_{\si \al} S_{\si \al}
        \ch^+_{\si\al}\Big)(h.c.)\notag
\\
   &\phantom{={}}{}+\Big(\sum_{\si \al} S^+_{\si \al}
   \ch^+_{-\si \al}\Big)(h.c.).
\end{align}
The squares can be {\it unfolded\/} by mean of an
integration over auxiliary gaussian fields. This amount to
introduce two real fields $a_0$ and $b_0$, one complex field
$a, \bar a$ and two pairs of Fermionic fields $\xi,\bar\xi$
and $\eta, \bar \eta$ with Gaussian measure 
\begin{align}
    d\mu(a_0,b_0,a,\bar a, \bar \xi , \xi,\bar\eta,\eta)
   &{}=e^{-V_0^2/2 -a_0^2/2 -b_0^2/2 -\vert a\vert^2/2-\xi^*
    \xi -\eta^* \eta}\notag
\\
   &\phantom{={}}{}\times\cdots\times\frac{dV_0\,da_0\,db_0
    \,d^2a}{2(2\pi)^{1/2}}\,d\xi^*\,d\xi\,d\eta^*\,d\eta.
\end{align}
Therefore
$$
    e^{-\la^2 (\cA + \cC)} = \int d\mu\,\exp\Big(i\la\sum_{\al}
        \Ph^+_{\al} R \Ph_{\al}\Big),
$$
with
\begin{align}
    \sum_{\al}\Ph^+_{\al} R \Ph_{\al}
   &{}=(a_0 + V_0 + \la^{-1} E)\sum_{\al,\si}
    S^{+}_{\si\al} S_{\si\al} \notag
\\&\phantom{={}}{}
   +(ib_0 + V_0 + \la^{-1} E)\sum_{\al, \si}
    \ch^{+}_{\si\al} \ch_{\si\al} \notag
   +\Big(\bar a \sum_{\al}S_{\al}S_{-\al}+  h.c.\Big)\notag
\\ &\phantom{={}}{}+\Big(\xi^*\sum_{\si\al}S^+_{\si\al}
   \ch_{\si\al}+ h.c.\Big)\notag
  +\Big(\eta^* \sum_{\si \al} S_{\si \al}
    \ch_{-\si \al} + h.c.\Big),
\end{align}
here $\Phi$ is the superfield $\Phi_{\alpha}^+ = (S^+_{\al},
S_{-\al}, \ch^+_{\al}, \imath\ch_{-\al})$ and $R$ is a
$4 \times 4$ supermatrix. Setting $A_0=(a_0+V_0+\la^{-1} E)$
and $\imath B_0 = (\imath b_0 + V_0 +\la^{-1} E)$ 
we obtain the following representation

\begin{equation}
\nu(E) = \frac{-1}{\pi}  \Im   \int 
d\mu    [\la R]^{-1}_{bb,11}  
\big[Sdet( \imath\la R)\big]^{-N}.
\end{equation} 
The matrix $R$ is a so-called 4x4 super-matrix
\be
\label{susy.eq-rmatrix}
R=\left(  \ba{cccc} A_0 &a &\xi^*&-\imath\eta           \\
    \bar a   &A_0  &\eta^*                 &-\imath\xi  \\  
 \xi &\eta  &\imath B_0   &0  \\
           -\imath\eta^*   &-\imath \xi^*   &0   &\imath B_0\\
\ea
 \right)\,=\,\left(      \ba{cc}          A                &\rho^{\ast}         \\
   \rho    &B  \ea   \right).
\ee
and the coefficient $ [\la R]^{-1}_{bb,11}  $ is a particular (bosonic)
coefficient of the inverse matrix, the one corresponding to first row and column.
In this computation we use the fact that this coefficient is also equal
to $ [\la R]^{-1}_{bb,22}  $ and that we get therefore $2N$ equal terms
from the sum over $\al$ and $\si$ in (\ref{susy.eq-supsersym1}), and cancel this sum with the
$1/2N$ normalization factor.

We perform a rescaling of all the variables by $\la^{-1}=\sqrt{2N}$ and write
\begin{eqnarray}
\nu(E) &=& \frac{-1}{\pi}  \Im \frac{\sqrt{N}}{2\sqrt{\pi}}  \int 
da_0 db_0 d a d\bar a  dV_0 d\xi^*\,d\xi\,d\eta^*\,d\eta \nonumber \\
&&e^{-N (a_0^2 + b_0^2 + |a|^2 + V_0^2 +2(\xi^*\xi+ \eta^*\eta))} \nonumber  \\
&& \bigl[  (A_0 - \frac{1}{iB_0}   (\xi^*\xi+ \eta^*\eta))^2  - ( a - \frac{2}{iB_0} \xi^*\eta  )
 ( \bar a - \frac{2}{iB_0} \eta^*\xi )  \bigr]^{-N-1} \nonumber \\
&& [A_0 - \frac{1}{iB_0}   (\xi^*\xi+ \eta^*\eta)] (iB_0)^{2N} , \label{expanfermio}
\end{eqnarray} 
defining $A_0 = a_0 + V_0 +  E $, $iB_0 = ib_0 + V_0+  E$.

We want to compute the fermionic integration 
\begin{eqnarray} I_f &=& \int d\xi^*\,d\xi\,d\eta^*\,d\eta \ e^{-2N(\xi^*\xi+ \eta^*\eta)   } 
\nonumber \\
&&\frac{[1 - \frac{1} {iB_0A_0} (\xi^*\xi+\eta^*\eta)]}
{\left[    [A_0 - \frac{1} {iB_0}  (\xi^*\xi+\eta^*\eta)  ]^2   - (a -  \frac{2} {iB_0} \xi^* \eta)
(\bar a -  \frac{2} {iB_0} \eta^* \xi)
\right]^{N+1} }. \label{fermionint}
\end{eqnarray} 

Note that
\be e^{-2N(\xi^*\xi+ \eta^*\eta)   } = 1 - 2N (\xi^*\xi+ \eta^*\eta) +4N^2 \xi^*\xi \eta^*\eta
\ee
\begin{eqnarray}
&&   \left[    [A_0 - \frac{1} {iB_0}  (\xi^*\xi+\eta^*\eta)  ]^2   - (a -  \frac{2} {iB_0} \xi^* \eta)
(\bar a -  \frac{2} {iB_0} \eta^* \xi)
\right]^{-N-1}  \nonumber\\
&=&   \left[    (A_0^2 - \vert a \vert ^2 )  -2 \frac{A_0} {iB_0}  (\xi^*\xi+ \eta^*\eta) \right.
\nonumber\\
&& \quad \quad \quad
+  \left. \frac{2} {iB_0}  (a\eta^* \xi  + \bar a \xi^* \eta) + \frac{6} {(iB_0)^2}   \xi^*\xi \eta^*\eta    
\right] ^{-N-1}  \\
&=&   \frac{1}{(A_0^2 - \vert a \vert ^2 )^{N+1}} \left[ 1 -
 \frac{N+1}{A_0 - \vert a \vert ^2 } 
\left(-2 \frac{A_0} {iB_0}  (\xi^*\xi+ \eta^*\eta) + \frac{2} {iB_0}  (a\eta^* \xi  + \bar a \xi^* \eta) 
\right.\right.
\nonumber \\
&& \quad\quad  + \left. \left.  
\frac{6} {(iB_0)^2}   \xi^*\xi \eta^*\eta  \right) 
+  \frac{(N+1)(N+2)}{A_0^2 - \vert a \vert ^2 }\frac{4} {(iB_0)^2}    \xi^*\xi \eta^*\eta  
\right]\ . \nonumber
\end{eqnarray} 

All terms vanish through fermionic integration except their terms in $ \xi^*\xi \eta^*\eta $.
So (\ref{fermionint}) can be computed as
\begin{eqnarray} I_f&=& \frac{1}{4\pi^2 (A_0^2 - \vert a \vert ^2 )^{N+1} }
\left[ \frac{N+1}{A_0^2 - \vert a \vert ^2 }  \frac{4N+2} {(iB_0)^2}  
- \frac{8N(N+1)}{A_0^2 - \vert a \vert ^2 }  \frac{A_0} {iB_0}  +4N^2\right.
\nonumber\\
&-& \left. \frac{1} {iA_0B_0}  \left(  \frac{N+1} {A_0^2 - \vert a \vert ^2}\frac{4A_0} {iB_0} 
-4N \right)
\right]\\
&=& \frac{1}{4\pi^2(A_0^2 - \vert a \vert ^2 )^{N+1} }
\left[ \frac{N+1}{A_0^2 - \vert a \vert ^2 } \left( \frac{4N-2} {(iB_0)^2}  -  \frac{8NA_0} {iB_0} \right) + 4N^2 + \frac{4N} {iA_0B_0} \right]\, . \nonumber
\end{eqnarray} 

We have therefore 
\be \nu(E) = \frac{-1}{\pi}  \Im \frac{\sqrt{N}}{2\sqrt{\pi}}  \int 
da_0 db_0 d a d\bar a  dV_0 e^{-N (a_0^2 + b_0^2 + |a|^2 + V_0^2)} (iB_0)^{2N} A_0 I_f\, ,
\ee
hence
\begin{eqnarray} 
\nu(E) &=& \frac{-\sqrt{N}} {4\pi^{7/2}}  \Im
 \int d\mu_b  (A_0^2 - \vert a \vert ^2 )^{-N-2}  (iB_0)^{2N-2} \\
&& \hspace{-1cm}
\left[(N+1)\left( (2N-1)A_0 -  4iNA_0^2 B_0 \right) + 
2iNB_0 (A_0^2 - \vert a \vert ^2 )(1+iNA_0 B_0) 
 \right] \ , \nonumber
\end{eqnarray} 
where
\be d\mu_b = da_0 db_0 d a d\bar a  dV_0 e^{-N (a_0^2 + b_0^2 + |a|^2 + V_0^2)} \ .
\ee
Expanding the Fermionic part and keeping only the leading terms in $N^2$
we would find
\begin{eqnarray}
\nu(E) &=& -\frac{N^{5/2}} {2\pi^{7/2}}  \Im
 \int d\mu_b  (A_0^2 - \vert a \vert ^2 )^{-N-2}  (iB_0)^{2N-2} \\
&& \hspace{-1cm}
A_0 \left[ 1 - 2iB_0 A_0 + (iB_0)^2 (A_0^2 - \vert a \vert ^2 )
 + O(1/N) \right] \ . \nonumber
\end{eqnarray} 
We use the notations $\cE  = -E/2 +i \sqrt{1-E^2/4}$
so that $\cE\cE^* =1$.

The ordinary saddle points correspond to $|a|=V_0=0$ and
for the main saddle point $a_0 =ib_0 = \cE $,
so that $A_0 = iB_0= -\cE ^*$
and $(iA_0 B_0)^{-1} = (\cE)^2 $.

At the second saddle point with same bosonic action 
 $a_0 = \cE, ib_0 = \cE^* $ we have  $iA_0 B_0 = 1 $ so that this contribution 
is $O(1/N)$.

Other saddle points have smaller actions \cite{BMR}.

The Hessian at the main saddle point is obtained by writing
 $a_0 \to\cE  + a_0$ and $ib_0 \to \cE +i b_0$ 
(which includes a translation of the integration 
contour) and expanding to second order.
The linear part vanishes at the saddle point and we find the Hessian approximation:
\begin{eqnarray}
\nu(E) &=& \hskip-.2cm \frac{N^{5/2}}  {2\pi^{7/2}}  \Im   \int 
da_0 db_0 d a d\bar a  dV_0 e^{-N \bigl[(\cE +a_0)^2 + (-i\cE + b_0)^2 
+ |a|^2 + V_0^2 )\bigr]}
\nonumber\\
&&e^{-N\bigl[\log [(\cE + a_0 + E+V_0)^2 - |a|^2] -2 \log (\cE +ib_0 +E +V_0) \bigr] } \nonumber\\
&& \cE  [   1 -2 \cE^2  +  \cE^4  ]   + O(1/N)\\
&=& \frac{N^{5/2}}  {2\pi^{7/2}}  \Im  \int 
da_0 db_0 d a d\bar a  dV_0 \nonumber\\
&&e^{-N \bigl[(a_0^2 + b_0^2 + |a|^2 )(1-\cE^2)+ V_0^2 
+2(ib_0 -a_0) V_0 \cE^2    \bigr] } \\
&& \cE  [   1 -\cE^2    ]^2   + O(1/N)\ .
\end{eqnarray} 
We can compute the Gaussian integral over $|a^2|$ as $\frac{2\pi}{N ( 1 -\cE^2 ) }$.
The remaining one is
$(\pi/N)^{3/2}$ times $1/\sqrt{{\rm det}\; Q}$ where $Q$ is the 3x3 matrix:
 \be
\label{eq-qmatrix}
Q =\left( \ba{ccc} 1-\cE^2 &0 &-  \cE^2  \\
0 &  1-\cE^2  & \imath \cE^2    \\  
 -\cE^2  &\imath \cE^2   & 1  \\
\ea \right)\ .
\ee
Since ${\rm det}\; Q = (1-\cE^2)^2$ we find:
\begin{equation}
\nu(E) = \frac{1}{\pi}   \Im  \cE =  \frac{1}{\pi} \sqrt{1-E^2/4}
\end{equation} 
which is the Wigner semi-circle law. 

This essentially ends the proof, up to two main effects which have been
neglected: we have to bound the error to the Hessian approximation to the saddle point and 
to treat the  "diagonal" $\cB$ quartic terms.

\subsection{Bounding the error to Hessian approximation}

First of all, we have to check that the path we have chosen
passing through the saddle is the correct one. This is true
if the real part of the action has minimum at the saddle.

Then we extract the leading contribution from the observable term,
(the corresponding integral is one by supersymmetry) that gives
the semicircle law. The remaining  error terms are of order $O(1/N)$.
To prove that we need to divide the integration region 
in the five parameters in four main zones
\begin{itemize}
\item[-] the vicinity of the ordinary saddle point: here the Hessian
approximation is correct. As we have extracted the leading term
already the numerator is small;
\item[-] the vicinity of the second saddle point: also here
the Hessian approximation is correct, but we have to expand
around the second saddle;
\item[-] a large but compact region around the two saddles: here we 
bound the integrand by supnorm;
\item[-]  the rest: as we are very far from the saddle we can
extract some exponential decay. This allows to perform
the integral and gives an exponentially small factor.
\end{itemize}

\medskip
\noindent{\bf Extracting the leading term}
\medskip

Putting together the terms of order $N^2$, $N$ and 1 we have
\be 
\nu(E) = \frac{-\sqrt{N}} {4\pi^{7/2}}  \Im
 \int d\mu'_b 
\frac{A_0}{(A_0^2 - \vert a \vert ^2 )} \left [N^2 I_1 \ + N   I_2 + I_3
\right ]
\ee
where the measure $\mu'_b$ contains also the logarithmic terms:
\be
d\mu'_b =  d\mu_b  \left (\frac{(iB_0)^2}{(A_0^2 - \vert a \vert ^2 )}\right )^{N} \ ,
\ee 
and
\begin{eqnarray} 
I_1 &=&  \frac{1}{(iB_0)^2(A_0^2 - \vert a \vert ^2 )} 2 \left [
1 - 2iB_0 A_0 + (iB_0)^2 (A_0^2 - \vert a \vert ^2 )
\right ] \; , \nonumber \\
I_2 &=&   \frac{1}{(iB_0)^2(A_0^2 - \vert a \vert ^2 )}  \left [
1 - 4 A_0 iB_0 + 2 \frac{iB_0}{A_0} (A_0^2 - \vert a \vert ^2 )\right ]\; ,
\nonumber \\
I_3 &=&  -\frac{1}{(iB_0)^2(A_0^2 - \vert a \vert ^2 )}  \ .
\end{eqnarray}

Note that, if we have no observable, the full supersymmetric integral is 1, and
its approximation to leading order is also 1! So we must have
\be
1 =  N^2 \frac{\sqrt{N}} {4\pi^{5/2}} 
 \int d\mu'_b  I_1 \ .
\ee
Therefore 
\be 
\nu(E) = + \frac{1}{\pi} \Im \cE + N^2 R_1 +N R_2 + R_3 \ ,
\ee
where 
\begin{eqnarray} 
R_1 &=&  \frac{-\sqrt{N}} {4\pi^{7/2}}  \Im  
\int d\mu'_b
\left [\frac{A_0}{(A_0^2 - \vert a \vert ^2 )} + \cE   \right ] I_1\; ,
\nonumber \\
R_2  &=&  \frac{-\sqrt{N}} {4\pi^{7/2}}  \Im  
\int d\mu'_b
\frac{A_0}{(A_0^2 - \vert a \vert ^2 )} I_2\; ,
\nonumber \\
R_3  &=&  \frac{-\sqrt{N}} {4\pi^{7/2}}  \Im  
\int d\mu'_b 
\frac{A_0}{(A_0^2 - \vert a \vert ^2 )} I_3 \ .
\end{eqnarray} 
We need to prove that the three quantities $N^2 |R_1|$,
$N |R_2|$ and $  |R_3|$ are  $O(1/N)$.

\medskip
\noindent{\bf Translation to the saddle}
\medskip

We translate to the saddle
\be
a_0 \to a_0 + \cE,\qquad b_0 \to b_0 -i \cE\ .
\ee
We do not cross any singularity (thanks to the $i \epsilon$ regulator)
hence the integral does not change.  The measure becomes 
\be
d\mu'_b =  da_0 db_0 d a d\bar a  dV_0 e^{-N S(a_0,b_0,V_0,|a|)}
\ee
and
\begin{eqnarray}
S &=&  \bigl[(\cE +a_0)^2 + (-i\cE + b_0)^2 + |a|^2 + V_0^2 )\bigr] 
\nonumber \\
&& + \bigl[\log [(-\cE^* + a_0 +V_0)^2 - |a|^2] -2 
\log (-\cE^* +ib_0 +V_0) \bigr] \nonumber \\
&=&  \bigl[ a_0^2 +  b_0^2 + 2 \cE(a_0-i b_0) + |a|^2 +V_0^2\bigr]
\nonumber \\
&&
+ \bigl[\log [(-\cE^* + a_0 +V_0)^2 - |a|^2] -2 
\log (-\cE^* +ib_0 +V_0) \bigr].
\end{eqnarray}

In order to bound the error terms, we will need a bound
on the absolute value of the measure.
So we have to study the critical points of Re $S$:
\be
{\rm Re} S = a^2_0 + b^2_0 + 2b_0\cE_i - 2a_0\cE_r + |a|^2 + V_0^2 
+ \frac{1}{2} \ln F_a - \ln F_b  \label{bigbla}
\ee
where we defined $\cE_i=\sqrt{1-E^2/4}$, $\cE_r=E/2$, 
\begin{eqnarray}
F_a &=& x^2 [x^2 + 2(\cE_i^2 -|a|^2)] + (\cE_i^2 +|a|^2)^2 \nonumber\\
F_b &=& (b_0+\cE_i)^2 + (V_0+\cE_r)^2  
\end{eqnarray}
and 
\be
x = a_0 + V_0 + \cE_r \ .
\ee
The equations for the critical points are
\begin{eqnarray}
\hspace{-.5cm}0 &=& \partial_{a_0}{\rm Re} S  =   2 \left \{ 
a_0 -\cE_r + \frac{x}{F_a} \left [ 
x^2 + \cE_i^2 - |a|^2 \right ]\right \} \; ,\nonumber \\
\hspace{-.5cm}0 &=& \partial_{b_0}{\rm Re} S  =   2 
(b_0 +\cE_i) \left [ 1 - \frac{1}{F_b} \right ]\; ,\nonumber \\
\hspace{-.5cm}0 &=& \partial_{a}{\rm Re} S  =   2 \bar a 
\left [
1 + \frac{1}{F_a} (-x^2 + \cE_i^2 + |a|^2 )
\right ] \; ,  \nonumber \\
\hspace{-.5cm}0 &=& \partial_{V_0}{\rm Re} S  =   2 \left \{ 
V_0 \left ( 1 - \frac{1}{F_b} \right ) -   \frac{1}{F_b} \cE_r+
 \frac{x}{F_a} \left [ 
x^2 + \cE_i^2 - |a|^2 \right ] \right \}. 
\end{eqnarray}
We remark that $E= O(\la^2) = O(1/N)$. Therefore we can
take $E=0$ in the saddle point computation, as it only
gives corrections of order  $O(1/N)$. At $E=0$ we have
$\cE_r=0$ and $\cE_i=1$. 
The saddle equations become
\begin{eqnarray}
\hspace{-.5cm} 0 &=& \partial_{a_0}{\rm Re} S  =   2 \left \{ 
a_0 + \frac{x}{F_a} \left ( 
x^2 + 1 - |a|^2 \right )\right \}\; , \nonumber \\
\hspace{-.5cm} 0 &=& \partial_{b_0}{\rm Re} S  =   2 
(b_0+1) \left [ 1 - \frac{1}{F_b} \right ]\; ,  \nonumber \\
\hspace{-.5cm} 0 &=& \partial_{a}{\rm Re} S  =   2 \bar a 
\left [
1 + \frac{1}{F_a} (-x^2 + 1 +   |a|^2 )
\right ]\; ,   \nonumber \\
\hspace{-.5cm} 0 &=& \partial_{V_0}{\rm Re} S  =   2 \left \{ 
V_0 \left ( 1 - \frac{1}{F_b} \right ) +
 \frac{x}{F_a} \left ( 
x^2 + 1 - |a|^2 \right ) \right \}\; , 
\end{eqnarray}
where now $x = a_0 + V_0 $,
\begin{eqnarray}
F_a &=& x^2 [x^2 + 2( 1 -|a|^2)] + ( 1 +|a|^2)^2 \; , \nonumber\\
F_b &=& (b_0+1)^2 + V_0^2  \ .
\end{eqnarray}
Note that, from the first and last equation we have
\be
V_0 \left ( 1 - \frac{1}{F_b} \right ) = - \frac{x}{F_a} \left ( 
x^2 + 1 - |a|^2 \right ) = a_0 \ . \label{409}
\ee
Moreover, from the third equation we have $a=0$ or 
 $(1-x^2+|a|^2) = -F_a$. In this last case we have
to solve 
\be
 x^2 [x^2 + 2( 1 -|a|^2)] + ( 1 +|a|^2)^2 =
-(1-x^2+|a|^2)\ .
\ee
We find there is no real solution for $y=x^2$. 
Therefore we have always $a=0$ in the following
and $F_a= (1+x^2)^2$. The saddle equations then reduce
to
\begin{eqnarray}
0  & = &    
a_0 + \frac{x}{x^2+1} \; , \nonumber \\
0  & = &  
(b_0+1) \left [ 1 - \frac{1}{F_b} \right ] \; ,\nonumber \\
0 & = &   
V_0 \left ( 1 - \frac{1}{F_b} \right ) +
 \frac{x}{1+x^2} \ .
\end{eqnarray}
 
Now we distinguish several cases. From 
the second equation we see that $b_0+1=0$ or $F_b=1$.

\medskip
\noindent{\bf First case: $F_b=1$}
\medskip

From (\ref{409}) we have $a_0=0$. We also have $x = V_0$, and
\be
\frac{x}{1+x^2}=\frac{V_0}{1+V^2_0}= 0 \ ,
\ee
which implies $V_0=0$. Finally $b_0$ is given by
\be
1= F_b = (b_0+1)^2 \qquad {\rm which \ implies \  \ } \ b_0=0, b_0=-2\ .
\ee
Therefore we have two saddle points: $a_0=V_0=a=0$ and $b_0=0,-2$.

\medskip
\noindent{\bf Second case: $b_0+1=0$} 
\medskip

In this case we have $F_b = V_0^2$, $a_0 = V_0 - 1/V_0$ 
and $x = (2V_0^2 -1)/V_0$. 
Inserting the expressions in terms of $V_0$ in the first equation we find
\be
\frac{V^2_0-1}{V_0} + \frac{V_0(2V^2_0-1)}{4V^4_0-3V^2_0+1} = 0
\ee
which gives the equation
\be
4 z^3 - 5 z^2 + 3 z -1 = 0
\ee
where $z=V_0^2$. 
Studying the first derivative of this expression we see
that there is only one real solution. This solution is positive
with $1/2 <z_s<1$. So there are two solutions $V_0=\pm \sqrt{z_s}$.
We have then two new critical points in $a=0$, $b_0=-1$, $V_0=\pm
\sqrt{z_s}$ and $a_0 = (V_0^2-1)/V_0$.

So finally we have identified four critical points:
\begin{eqnarray}
S_1 & \equiv &  a_0=V_0=a=b_0=0  \; ,\nonumber\\
S_2 & \equiv &  a_0=V_0=a=0,\ b_0= -2 \; ,\nonumber\\
S_3 & \equiv &  a=0,\ b_0= -1,\ a_0 =  \frac{(z_s-1)}{\sqrt{z_s}}, \ 
V_0=\sqrt{z_s} \; ,\nonumber\\
S_4 & \equiv & a=0,\ b_0= -1,\ a_0 = - \frac{(z_s-1)}{\sqrt{z_s}}, \ 
V_0=-\sqrt{z_s} \ . 
\end{eqnarray}
The real part of the action at the leading order ($E=0$) is
\be
{\rm Re}_{E=0} S = a^2_0 + b^2_0 + 2b_0  + |a|^2 + V_0^2 
+ \frac{1}{2} \ln F_a - \ln F_b \ .
\ee
Is is easy to see that in $S_1$ and $S_2$ we have ${\rm Re}_{E=0} S=0$.
On the other hand in $S_3,S_4$ the action is (writing $z$ for $z_s$):
\be {\rm Re}_{E=0} S = \frac{(2 z -1)(z-1)}{z} + \log \frac{4z^2 - z +1}{z^2}
\ee
which since $1/2 < z <1$ satisfies
\be {\rm Re}_{E=0} S > -1/4   + \log (3/2) > 0.15 \ .  
\ee

\medskip
\noindent{\bf Integration regions}
\medskip

We call $X=(a_0,b_0,V_0,a)$ so that the integral to
study is
\be
\int dX\ e^{-N{\rm Re}S(X)} |R(X)|
\ee
where $R(X)$  depends on which error term we are looking at.
We  introduce the norm 
\be
\Vert X \Vert^2 = X^* X = a_0^2 + b_0^2 + |a|^2 + V_0^2\ .
\ee
Now we partition the integration domain in four regions
$T_i$, $i=1,...,4$, where
\begin{eqnarray}
T_1 &=&  \left \{ X | \ \Vert X\Vert \leq \frac{1}{N^{1/3+\de}}  
\right\} \; ,\nonumber\\
T_2 &=&  \left \{ X | \ \Vert X-S_2\Vert \leq \frac{1}{N^{1/3+\de}}
\right \} \; , \nonumber\\
T_3 &=&  \left \{ X | \ \Vert X\Vert < K\ {\rm and}\ X\not\in T_1\cup
T_2   \right \}\; ,  \nonumber\\
T_4 &=&  \left \{ X | \ \Vert X\Vert > K  \right \}\; ,
\end{eqnarray}
where $0<\de<<1$ and $K>>1$ are fixed constants.

\medskip
\noindent{\bf Bound in the region $T_1$}
\medskip

In this region we can expand the action to the second order.
We get
\begin{eqnarray}
N{\rm Re} S &=& N \bigl[(a_0^2 + b_0^2 + |a|^2) 2 \cE_i^2+ V_0^2 
 + 2[b_0  2 \cE_r\cE_i +  a_0 
(\cE_i^2-\cE_r^2)]V_0   
\nonumber\\
&& \quad \quad \quad \quad  \quad \quad \quad \quad \quad \quad \quad \quad+ O(\Vert X\Vert^3)  \bigr]
\nonumber \\
 &=& N X^* H X + O(N^{-3\de}) 
\end{eqnarray}
where
\be
H =\left(
\ba{cccc}
2\cE_i^2         & 0            & \cE_i^2 -\cE_r^2 & 0 \\
0                &2\cE_i^2      & 2 \cE_r\cE_i     & 0\\
\cE_i^2 -\cE_r^2 & 2 \cE_r\cE_i & 1                & 0\\
0                &  0           &  0               & 2\cE_i^2 \\       
\ea
\right ) = 
\left(
\ba{cc}
{\rm Re} Q & 0\\
 0         &  2\cE_i^2\\
\ea
\right )
\ee
where $Q$ was defined in (\ref{eq-qmatrix}).

Note that $H$ is positive definite, so the measure is well
defined.
Now we remark that in this region
\begin{eqnarray} 
I_1 &=&  (1-\cE^2)^2 + O(\Vert X\Vert) \; ,\nonumber \\
I_2 &=&  \cE^2 (\cE^2-2)  + O(\Vert X\Vert) \; ,\nonumber \\
I_3 &=&  -\cE^4  + O(\Vert X\Vert) \; ,
\end{eqnarray} 
and 
\be
\left [\frac{A_0}{(A_0^2 - \vert a \vert ^2 )} + \cE   \right ] =
 O(\Vert X\Vert) \ .
\ee
Inserting this bounds in $R_1$,$R_2$, $R_3$ we get
\begin{eqnarray} 
N^2|R_1|_{T_1} &\leq&  const \  N^{5/2}  
\int dX e^{-NX^* H X} O(\Vert X\Vert) = O \left (\frac{1}{\sqrt{N}} \right )\; ,
\nonumber \\
N|R_2|_{T_1}  &\leq &    const \   N^{3/2}    
\int  dX e^{-NX^* H X} = O\left (\frac{1}{N} \right )\; ,
\nonumber \\
|R_3|_{T_1}  &\leq & const  \sqrt{N}    
\int dX e^{-NX^* H X}  = O\left (\frac{1}{N^2} \right ) \ .
\end{eqnarray} 
By parity we can improve the first estimate and obtain in fact rather easily $N^2R_1 =O(1/N)$.

\medskip
\noindent{\bf Bound in the region $T_2$}
\medskip

In this region, we expand around the second saddle $S_2$ and
we obtain
\begin{eqnarray}
N{\rm Re} S_{T_2} &=& N \bigl[(a_0^2 + (b_0+2)^2 + |a|^2) 2 \cE_i^2+
V_0^2 \nonumber \\
 &&+ 2[(b_0+2)  2 \cE_r\cE_i +  a_0 
(\cE_i^2-\cE_r^2)]V_0   \bigr] + O(\Vert X-S_2\Vert^3)N\nonumber \\
 &=& N (X-S_2)^* H (X-S_2) + O(N^{-\de}) \ . 
\end{eqnarray}
The bounds on $NR_2$ and $R_3$ work as before. For $N^2R_1$ we have
\be 
I_1 =  0 + O(\Vert X- S_2\Vert) \ ,
\ee
\be
\left [\frac{A_0}{(A_0^2 - \vert a \vert ^2 )} + \cE   \right ] =
 O(\Vert X- S_2 \Vert) \ ,
\ee
so that again $N^2R_1 = O(1/N)$.

\medskip
\noindent{\bf Bound in the region $T_3$}
\medskip

We have to bound  $f(X) = | e^{-N Re S(X)}  R(X)|$.
We shall prove
\be f(X) = | e^{-N Re S(X)}  R(X)| \le c e^{-N^{1/3 -3\de}}\ .
\ee
The function $R(X)$ is a rational fraction and to bound it
we must take care of the possible zeroes of the denominator.
This denominator is $(A_0^2 - |a|^2)^2 (iB_0)^2$. 
But $(A_0^2 - |a|^2)$ has no zeroes thanks
to the imaginary part $\cE_i$. $iB_0$ in contrast  vanishes on the submanifold
$b_0 = -\cE_i$ and $V_0 = - \cE_r$., but the factor $(iB_0)^{2N}$ that
we put in $S$ also vanishes. So we are lead to write 
\be e^{-N S(X)}  R(X)  = e^{-(N-2) S(X)  -2 S_2(X) + \ln G_a}  R'(X)
\ee
where 
\be G_a = (-\cE^* + a_0 +V_0)^2 -|a|^2
\ee
$R'(X) = R(X) (iB_0)^2$ is now a rational fraction with a non-vanishing
denominator, and
\be S_2 (X) = a_0^2 + b_0^2 + 2\cE (a_0 - ib_0) + |a|^2 + V_0^2\ .
\ee
On the compact region $K$ we have 
\be  e^{ -2 S_2(X) + \ln G_a}  \le K_1
\ee
for some constant $K_1$, and
\be  |R'(x)|  \le c N^{5/2}(1+ \Vert X\Vert )^{d} 
\label{glouglou}
\ee
for some constants $c$ and $d$.

So it is sufficient to prove that in that region
\be
(N-2){\rm Re} S_{T_3} \geq  c (N-2) \frac{1}{N^{2/3+2\de}} = c'   N^{1/3-2\de} \label{bloublou}
\ee
for some small positive constant $c$ and $c'$. 
Indeed we have to check where ${\rm Re} S$ can reach its minimum. This can only be on a critical point or on the boundary. But the only critical points in $T_3$ are $S_3$ and $S_4$, where
we have the much stronger bound $N{\rm Re} S   \ge 0.15 N $. On the boundary with $T_1$
and $T_2$ the previous Hessians approximations are still valid and give
the bound (\ref{bloublou}). Finally on the outer boundary the bound is even better (see below). 

\medskip
\noindent{\bf Bound in the region $T_4$}
\medskip

When we are far enough from all saddle points we can apply a
rough bound to the logarithms in the action
\be
e^{N[\frac{1}{2}\ln F_a -(N-2)\ln F_b]} \leq e^{+  K'N (\log \Vert X\Vert) }  
\ee
for $K'$ big. Moreover 
\be
 e^{+  K'N (\log \Vert X\Vert) }  \le e^{+ N \Vert X\Vert^2 /4}
\ee
taking $K$ (the parameter defining region $T_4$) large enough,
since recall that in $T_4$ $\Vert X\Vert \ge K$.

Finally in $T_4$ it is easy to check that for large enough $K$
\be
(N -2) (a_0^2 + b_0^2 c+ 2b_0 \cE_i -2 a_0 \cE_r + |a|^2 + V_0^2) \geq  (1/2) N   \Vert X\Vert^2 
\ee
so that remembering (\ref{bigbla}) and the bound (\ref{glouglou}) on $R'$ we can bound 
\be
\int_{T_4} dX  e^{-N {\rm Re} S(X)} |R(X)|  \le \int_{\Vert X \Vert >K}  e^{- N Vert X\Vert^2 /4} N^{5/2}
K_2 (1+ \Vert X\Vert )^d \le e^{-cN} \ .
\ee
This completes the proof.

\subsection{Bounding the correction due to $\cB$}

This is explained in a rather detailed 
way in the last section 6, of \cite {BMR}, 
which for completeness we roughly reproduce here.

The quartic term $\cB$ should be small as $1/N$ compared to the other ones
since it contains only one sum over sectors $\alpha$, not two
independent sums over $\alpha$ and $\beta$. However it
has the wrong sign, so although statistically small, it is dangerous
at large fields. The solution is to treat it by some kind
of small/large field expansion. A Taylor expansion with 
integral remainder is written successively for
each of the $N$ sectors appearing in the sum for $\cB$:
\be
    \cB = \sum_{\al} \cB_{\al}.
\ee
To first order the expansion gives
\be
    e^{-\la^2 \cB_{\al}} = 1 - \int\limits_{0}^{1}
   \la^2 \cB_{\al} e^{-\la^2\cB_{\al}} e^{+t\la^2\cB_{\al}}\,dt.
\ee
This Taylor expansion either suppresses $\cB_{\al}$ from the
exponential of the action or generates a remainder term
\be
    R_{\al}= -\int\limits_{0}^{1}\la^2\cB_{\al} e^{-\la^2\cB_{\al}}
        e^{+t\la^2\cB_{\al}}\,dt.
\ee
The rough line of argument is as follows: either the set $P$
of sectors where $\cB_{\al}$ is not suppressed is small or it is large.
Let $p=\vert P\vert$. If $p$ is large, we have gained such a small probabilistic factor that we
can treat this case by a rough translation of the $V$
contour, like for large field cubes of section (3.3) (regime $|p^2 -E|>> \la^2$).
No saddle point analysis is necessary.

If $p$ is small, we have to perform the saddle point analysis, but we
can restrict is to the sum of sectors outside $P$ where it is essentially
the same analysis than previously but with only $N'=N-p$ sectors.
We have to treat also the coupling between the sectors in and outside $P$,
and this is a delicate point, since the ultimate reason for which these couplings are small
is the supersymmetry of the underlying field theory.

Now what "small" or "large" means for $p$? It means $p$
smaller or larger than some function $p(N)$. Since we expect a small factor
of order $1/N$ for each sector in $P$ (through the lacking sum over $\beta$),
and a rough imaginary translation on $V$ costs $K^N$,  we need
$K^N e^{-p(N) \log N} << 1$. 

Therefore we choose for $p(N)$ the integer part of $N/\sqrt{\log N}$. The
expansion is stopped at order $p(N)=p$. This means that we write:
\be
    e^{-\la^2\cB}=1+\mathop{\sum}
\limits{P\subset[1,...,N]\\ 
0<|P|\le p(N)}                 
R_P,
\ee
where
\begin{alignat}{2}
    R_{P}
   &{}= \prod_{\al \in P} R_{\al}
   &\qquad
   &{\rm if}\;\;|P|<p(N),
    \notag
\\
\label{susy.eq-rpterm}
    R_{P}
   &{}=\prod_{\al\in P}R_{\al}\prod_{\al>\max P}
    e^{-\la^2\cB_{\al}}
   &\qquad
   &{\rm if} \;\;|P| = p(N).
\end{alignat}
Let $Q$ be the complement of $P$ in $[1,...,N]$. The term
$1$ was treated in the previous sections. The remainders
terms $R_P$ must be shown to be $O(1/N)$ as $N\to\infty$.
Let $R_{p(N) } = \mathop{\sum}\limits{P \subset [1,...,N] \\ |P| =p(N)} R_{P}\Mf$ 
be considered first. For this term we said that it is not
necessary to perform any saddle point analysis. It is
sufficient to return to the treatment of subsection 3.3, so we undo
the $V$ integration: all the
$e^{-\la^2\cB_{\al}}$-terms are recombined with the $\cA$-term to
reproduce the initial functional integral~%
(\ref{susy.eq-supsersym1}).
However we cannot exactly reproduce the initial functional
integral over $V$, since some $e^{-\la^2\cB_{\al}}$-factors are missing
or appear with reduced weights in (\ref{susy.eq-rpterm}). This means that there
remains quartic
correction terms $e^{t\la^2\cB_{\al}}$ or $e^{\la^2\cB_{\al}}$. The
important remark is that the bosonic part of these terms has
now the right sign~! Therefore they can be represented as a
well defined functional integral over a new auxiliary field $W_{\al}$.
For instance
\begin{align*}
    \exp\big(-2\la^2 t S^+_{\al }S_{-\al}S^+_{-\al }S_{\al}\big)
   &{}=\exp\big(-2\la^2 t \vert S_{\al } S_{-\al }\vert ^2\big)
\\
   &{}=\int dW_{\al}\,d\bar W_{\al}\,e^{-\vert
    W_{\al}\vert^2}\exp\big(\imath \la \sqrt{2t} W_{\al} S_{\al}
    \bar S_{-\al } + cc\big).
\end{align*}
With slightly condensed notations, this leads to
\begin{eqnarray}
R_{p(N)} &=&
\sum_{P\subset[1,...,N]\  , \  |P| = p(N)}
\int S^+S\Big(\prod_{\al \in P}\int 
\limits_{0}^{1}
        \la^2 \cB_{\al}\,dt\Big) \nonumber\\
&=&\times\exp\big( i\Psi^+(E+ \la V+ \la \sqrt{2t}W) \Psi\big)\,d\Psi^+ \,d\Psi \,d\mu (V, W).
\end{eqnarray}
Then a complex translation $V_0 \mapsto V_0 \pm i
\la^{-1}$ is performed, with the same sign as the imaginary
part of $E$ in order to avoid crossing of singularities. In
other words
\be
    \int\limits_{-\infty}^{+\infty}e^{-V_0^{2}}F(V_0)\,dV_0=
    \int\limits_{-\infty}^{+\infty} e^{-V_0^{2}}
    e^{-2i\la^{-1}V_0^{2}}e^{\la^{-2}}
        F(V_0+i\la^{-1})\,dV_0.
\ee
The functional integral can now be bounded by its absolute
values everywhere, namely the following contributions are
bounded
\begin{itemize}
\item[-]
    by $2^N$, for the sum over $P$, that is the total
    number of subsets of $[1,N]$;
\item[-]
    by $1$, for the integrals such as $\int_{0}^{1} dt$;
\item[-]
    by 1, for the oscillating imaginary integrals;
\item[-]
    by Gram's inequality for fermions or the Schwarz
    inequality for Bosons, for the remainders terms
        $\cB_{\al}$;
\item[-]
    by 1, for every propagator since the imaginary
        translation in $V_0$ has created an imaginary part
        proportional to the identity in the denominator of the
        Green's function.
\end{itemize}
This means that each $\cB_{\al}$ term gives rise as expected to a small
factor $\la^{2}=1/2N$ for each sector $\alpha$ in $P$, hence 
altogether we have a small factor
$1/N^{p(N)}$. The two source terms are bounded by 1. The
normalization determinants are then easily bounded by
$c^{N}$, even without using the supersymmetry cancellations,
since the operators considered are bounded in a finite $2N$
dimensional space thanks to the imaginary part of $E$ which
is no longer infinitesimal. Combining all factors leads to
\be
    R_{p(N) }\le c^{N}  e^{-c N \sqrt{\log N}},
\ee
showing that this correction term is indeed small.

It remains to treat the terms with $1\le \vert P \vert\le p(N)$. 
To bound these terms a mean-field analysis will be performed 
like in subsection (3.2), introducing the $a$ and $b$ 
fields, but we said it should apply only to the sectors of the theory in the 
complement $Q$ of $P$. The functional integral to be bounded 
for a single term is~(\ref{susy.eq-rpterm}). Now, the 
$e^{-\la^2 \cB_{\al}}$-terms are recombined only for $\al \in P$ 
with the $\cA$ term to reproduce the initial functional 
integrals over the $V$ fields and the terms 
$e^{t\la^2 \cB_{\al}}$, also for $\al \in P$, are again given by 
defined integrals over new auxiliary fields $W_{\al}$. 
Finally, the quartic terms, with sector sums reduced to $Q$, 
are treated exactly as in the previous section, hence
mean-fields $a, a_0, b_0$  are correspondingly introduced. 
This leads to a representation
\be
    R_P=\int dV_{P,P}\, dV_{P,Q}\, d^2a\,dV_{0}\,da_0\, db_0
    \prod_{\al \in P} R_{\al}e^{\cL_{Q}},
\ee
where $Q$ is the complement of $P$ in $[1,...,N]$. In 
addition, $V_{P,P}$ is the part of the matrix $V$ 
corresponding to rows and columns in $P$, including the new 
fields of the $W$ type. $V_{P,Q}$ correspond to one entry in 
$P$ and the other in $Q$, and the mean field computation is 
now restricted to $Q$. The integral over superfields gives
rise again to a superdeterminant and an 
additional correction, whose bosonic part (leaving the Fermionic part 
to the reader) is of the type

\be
    \exp{\big[Tr\log (1+ C V)\big]},
\ee
where $C$  is the bosonic part of the $R ^{-1}$ matrix for the $N-p$ analogous 
problem, as in the previous subsection 
(see eq.~(\ref{susy.eq-rmatrix})), and $V= V_{PP} + V_{P,Q}$ 
is the perturbation. This correction term is bounded by
\be
    \big|\exp{[Tr\log (1+ K)]}\big|\le
        \exp{(Tr K+K^{*} + KK^{*})}.
\ee
Evaluating $C= R ^{-1}$ at the saddle point costs a factor
$\exp{(N p/2N)} = e^{2p}$ at most. Each term $\cB_{\al}$ 
naively gives a factor $\la^{2}=1/(2N)$ when evaluated, but 
this simply compensates the sum over $\al$ when $p$ is 
nonzero but small, so we have to gain an additional $1/N$ 
factor. Adding a few expansion steps gives such a small 
additional factor $1/N$ (already for the first non zero 
value $p=1$). This can be seen by integrating by parts the
superfields in the {\it vertex\/} $ \cB_{\al}$ which has
been taken down the exponential. By supersymmetry, the
{\it vacuum graph\/} corresponding to a contraction of the
four fields at the vertex vanishes. This is absolutely
necessary since this graph by simple scaling is proportional
$1/N$ and cannot have any additional $1/N$ factor. Its
vanishing can be checked by hand:
\begin{itemize}
\item[-] 
    the self-contractions of the bosonic piece $2 S^+_{\al }
        S_{-\al}S^+_{-\al }S_{\al}$ give a factor +2;
\item[-]
    the self-contractions of the boson-fermion piece
        $$
            \Big(\sum_{\si}S_{\si\al}^{+} \ch_{-\si\al}^{+}\Big)
                \Big(\sum_{\si '} S_{\si\al} \ch_{-\si\al}\Big),
        $$
    give a factor -2 (since there is one Fermionic loop
        giving the minus sign and one sum over $\si$ giving a
        factor 2 only after the contractions);
\item[-]
    the selfcontractions of the term $(1/2)\sum_{\si}
        ( \Psi^{+}_{\si \al}\Psi_{\si \al })^2$ are clearly
        supersymmetric and also add up to 0.
\end{itemize}
Consequently, at least one field of the {\it vertex\/}
$\cB_{\al}$ has to contract to the exponential. Performing
two contractions in turn, gives always at least a factor
$p/N^2$ at the end instead of the naive $1/N$ factor. Indeed
the worst case corresponds to the non trivial contraction
term being of the type $V_{PQ}$. This generates a new factor
$1/N$ but a new sum over $\beta \in Q$, which costs $N-p
\simeq N$ so nothing is gained yet. But contracting this
$\beta$ field again either generates a diagonal term, hence
a new $1/N$ factor, with no new sum, or returns to a
$V_{Q,P}$ term. This last situation generates a new $1/N$
factor and a new sum over sectors $\gamma$ but this time
this new sum is restricted to $P$, so it costs only a factor
$p$ instead of $N$! Hence at worst, after these two
contraction steps, a total factor $p/N^2$ instead of the
naive factor $1/N$ is associated to each vertex, as
announced. Now the sum over~$P$ costs a total factor
$N!/p!(N-p)!$, hence is bounded by $c^{p} [N/p]^{p}$.
Combining all factors, the sum of contributions of such
terms with $p\ne 0$ is bounded by $\sum_{p=1}^{+\infty}
[c/N]^{p} \le c'/N$. It is therefore at least as small
as $1/N$.

This completes our sketch of the content of \cite{BMR}. 
The last section in this review
is devoted to generalizations that better mimic the symmetries of the 
2-dimensional
matrix model near degeneracy of the rhombus, using anisotropic sectors.

\section{Improved Flip Matrix Models}

As we already said in Section 3.2, the isotropic sector
model is not optimal for momentum conservation analysis.

Actually we know that the potential $V(\gamma)$, has to be associated
to pairs of vectors  $p,q$ with
$|p|=1+O(M^{-j})$, $|q|=1+O(M^{-j})$ (using the same
notation as in previous sections)  and $p-q=\gamma$. 
By simple geometric arguments we can see that the angle
$\sigma$ between $p$ and $q$ is then 
$\sin (\si/2)\simeq  |\gamma|/2$  and $p$ can move of an
angle $\de \si = O(M^{-j}/|\gamma|)$. 

\begin{itemize}

\item[-] If we have isotropic sectors (of width $M^{-j}$), then
for each  $|\gamma|$ there is a set $S(\gamma)$ of $O(1/|\gamma|)$ 
pairs of sectors $\al,\bt$ (with angle $\si$ between them) associated to the 
random variable $V(\gamma)$. So when $|\gamma|<1$ the matrix
elements $V_{\al\bt}$ for $(\al,\bt)\in S(\gamma)$ are no longer
independent random variables.
In the limit case, $|\gamma| = O(M^{-j})$ all the 
$O(M^{j})$ sectors are associated to the same $V(\gamma)$. This
is the diagonal term  $V_0$ that appears in the Isotropic flip model.

\item[-] If we still want to work with matrix elements which
are all independent random variables we have to introduce anisotropic
sectors of width $M^{-j/2}$. In this case the matrix elements
are all independent for $\gamma\geq M^{-j/2}$. The case $\gamma<M^{-j/2}$
corresponds to the diagonal terms $V_{\al\al}$. Here we do expect
to have no longer independence, but the problem arises only on the
diagonal terms, just as in the Isotropic flip model.
The drawback is that now, for each sector $\al$ 
we have also to introduce a dual space decomposition $R_\al$ of the
spatial cube $\De$ of side $M^j$ in  rectangles of side $M^j$
in the direction parallel to $\al$ and $M^{j/2}$ in the orthogonal
direction. The random matrix is then indexed by  pairs $(\al,t)$
where $\al$ is a sector and $t$ a dual rectangle.

\end{itemize}

\subsection{The Anisotropic (2d) Flip Matrix Model}

As in the previous sections we consider the Anderson model
in a spatial cube $\De$ of side $M^j$, and we introduce a cut-off
in momentum space that selects only an annulus of width $M^{-j}$ 
around the Fermi surface. We fix then  $M^j=\la^{-2} = 2N$, $E_0=1$
and $E=O(\la^2)=O(N^{-1})$.

We then cut the annulus  in anisotropic sectors of
width $M^{-j/2}$, so that a sector now has width $M^{-j}$
in the radial direction and $M^{-j/2}$ in the tangential
direction. Therefore we have $M^{j/2}$ anisotropic
sectors instead than $M^j$ isotropic ones. 
For each sector $\al$ then we introduce a partition $R_\al$
of the cube $\De$ in rectangles of side $M^j$
in the direction parallel to $\al$ and $M^{j/2}$ in the orthogonal
direction.  

In this way we obtain a finite dimensional space of 
dimension $2N=M^j$ with a new index for its orthogonal basis, 
namely the pairs $(\alpha, r,\sigma)$ where $\alpha$
runs over the first $\sqrt{N}$ anisotropic directions, 
$r \in R_{\alpha}$ runs over the
$\sqrt N$ rectangles elements of $R_{\alpha}$ 
and as previously $\sigma$ takes two values $\pm 1$ 
corresponding to the two opposite sectors
on the circle. The dimension of the Hilbert space for the matrix
model we want to construct is then the same as in the isotropic case.

We list now the columns and rows of our random matrix as before,
with sectors from 1 to $ \sqrt N$, then from $-\sqrt N $ to -1, and 
for each sector $\alpha$ we list the new rectangular variables 
$r$ in direct order according to an arbitrary ordering of
$R_{\alpha}$. Note that the rectangles for $\al$ and $-\al$ are then
the same modulo a rotation of $\pi$. The effect of the
rotation is that the ordering of the rectangle is inverted (see Fig.\ref{ord}).
This is actually good as it ensures the symmetry of the matrix around
the anti-diagonal.

Finally, as in the case of the isotropic sectors,
we discretize the potential and build a random matrix model,
that will be called Anisotropic Flip Matrix Model (AFMM).

\subsubsection{Construction of the matrix model}

To construct the matrix we introduce a discretization of
the phase space for $V$, in such a way to ensure that:
\begin{itemize}
\item[-] the off-diagonal matrix elements are independent random
         variables with covariance one. 
\item[-] the dependence among the  diagonal terms is explicitely extracted.
This means each $V_{\al\al}$ is a sum of independent 
random variables $V_{\al\al}=V_0 + V_1 + ...+ V_{j/2}$ where 
$V_0$ appears in all diagonal terms and $V_{j/2}$ only in one.
\end{itemize}

Note that as $V$ is restricted to act on the shell $|p^2-1|=O(M^{-j})$,
the momentum $\gamma$ for $V$ is restricted to $|\gamma|\leq 2 + O(M^{-j})$.
In position space $V$ is restricted to the cube $\De$.

To construct  the off-diagonal matrix elements we divide
the momentum region $\gamma$ in three parts:
\begin{itemize}
\item{} a {\em  forward-momentum} region
corresponding to $M^{-j/2}< |\gamma|<1$ (sector pairs at an angle
$M^{-j/2}< \si<\pi/3$),
\item{} a {\em backward momentum}  region
corresponding to  $1\leq |\gamma|\leq 2 + O(M^{-j})$
(sector pairs at an angle $\pi/3\leq \si \leq \pi + O(M^{-j})$),
\item{} a {\em 0-momentum} region corresponding to   $|p| \le M^{-j/2}$
($\si\leq M^{-j/2}$); this last region gives the diagonal terms of the matrix
\end{itemize}

\begin{figure}
\medskip
\medskip
\centerline{\psfig{figure=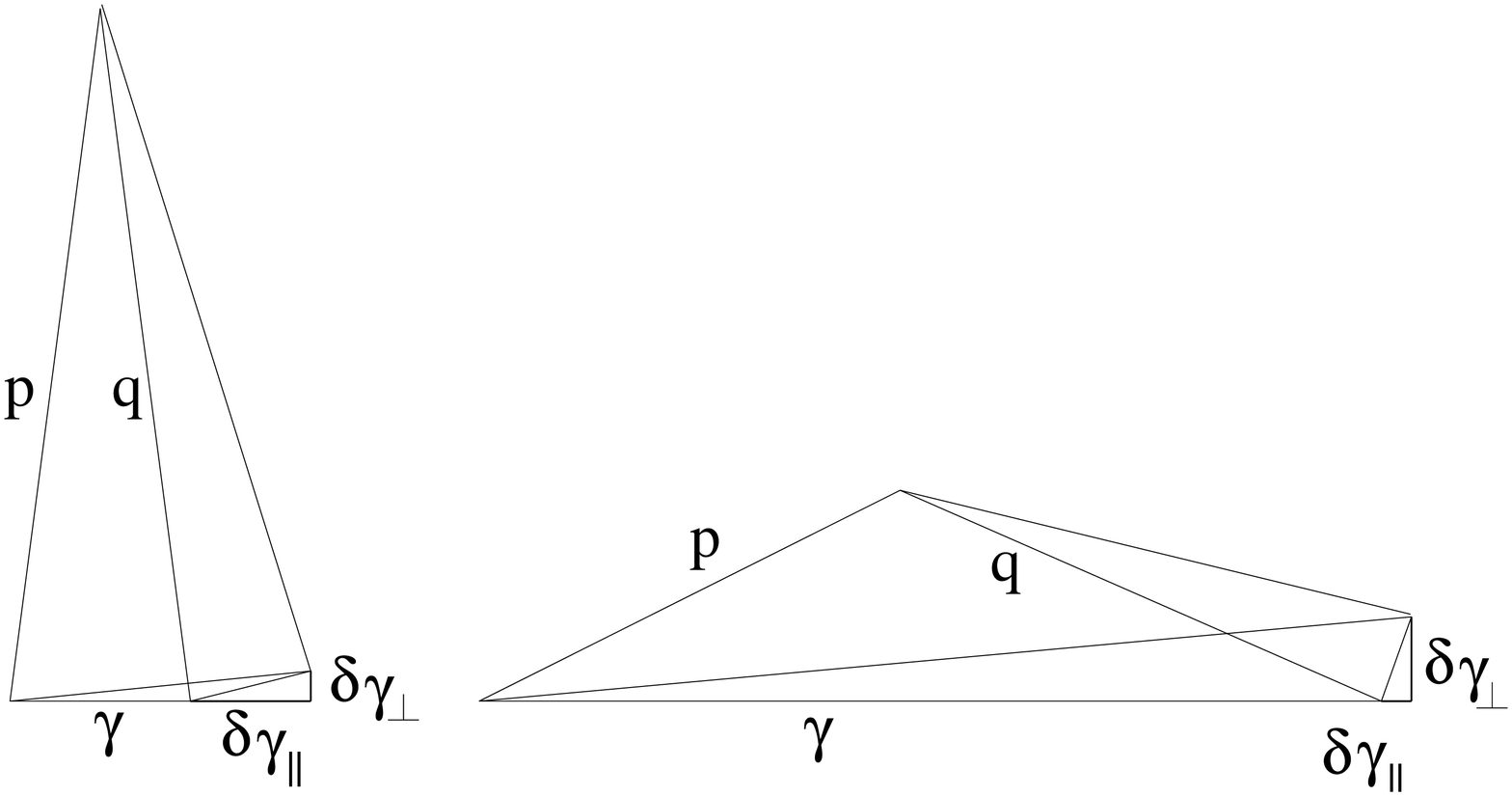,width=7cm}}
\caption{Variation of $\gamma$ as $p$ and $q$ move inside their sectors}
\label{gamma}
\end{figure}

\begin{figure}
\centerline{\psfig{figure=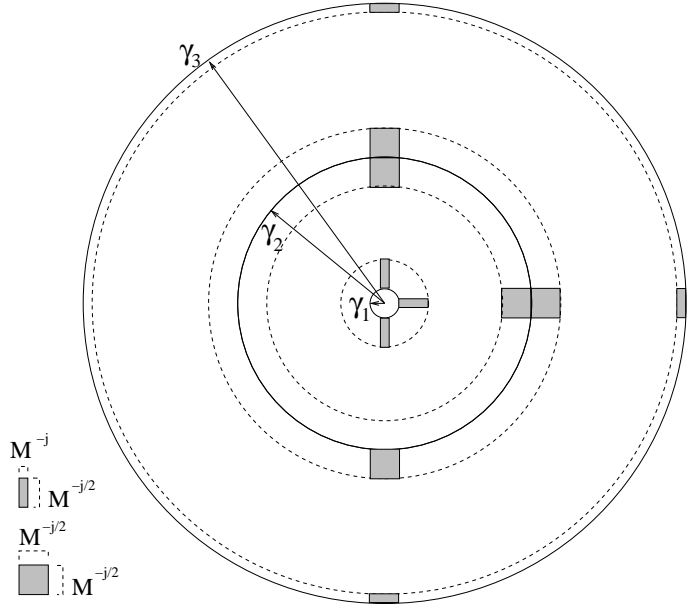,width=9cm}}
\caption{Different decomposition depending on $|\gamma|$: for 
$|\gamma|=|\gamma_3|\simeq 2$ we have to use rectangles of side $M^{-j}$ in one
direction and $M^{-j/2}$ in the other direction, for
$|\gamma|=|\gamma_2|\simeq 1$ we have to use cubes of side  $M^{-j/2}$ and
for $|\gamma|=|\gamma_1|\simeq M^{-j/2}$ we have to use again rectangles,
but exchanging the sides}
\label{shells}
\end{figure}

In these  regions the discretization of the phase space for $V$
must be different. 
To see that let us consider a pair of sectors $\al,\bt$
with two vectors inside $p\in\al,q\in\bt$ and  $\gamma=\al-\bt$.
As $p$ and $q$ move inside their sectors the variation for $\gamma$
is given by 
\be
\de\gamma_{\Vert} = M^{-j/2}\ \cos(\frac{\si}{2}+ M^{-j/2})\qquad
\de\gamma_{\perp} =  M^{-j/2}\ \sin(\frac{\si}{2}+ M^{-j/2})\ .
\ee
as $M^{-j/2}$ is the sector size.
Now we distinguish four limit situations (see Fig.\ref{gamma}-\ref{shells})

\paragraph{1.} When $|\gamma|\simeq 1$ ($\si\simeq \pi/3$) then 
$\de\gamma_{\Vert}\simeq \de\gamma_{\perp} =  O(M^{-j/2})$
so we should cut momentum space in cubes of side $M^{-j/2}$.
This corresponds to $\ga_2$ in Fig.\ref{shells}.

\paragraph{2.}  When $|\gamma|\simeq M^{-j/2}$ 
($\si= O(M^{-j/2})$) we are almost on the diagonal part of the matrix 
($\ga_1$ in  Fig.\ref{shells}). The value of
$\gamma$ is fixed with
a precision much higher than 
$M^{-j/2}$ in the  perpendicular direction, that is 
$\de\gamma_{\Vert}=  O(M^{-j/2})$ while 
$\de\gamma_{\perp} =  O(M^{-j})$. In this region
we must then use rectangles of width $M^{-j/2}$ in the radial
direction (that of $\gamma_{\Vert}$) and $M^{-j}$ in the 
orthogonal direction. 

\paragraph{3.}  
When $|\gamma|= 2-M^{-j}$ ($\si\simeq \pi$) the situation is reversed.
The vector $\gamma$ is fixed with high precision in the 
parallel direction. For  $\de\gamma_{\Vert}$ we have
\be
\de\gamma_{\Vert}=M^{-j/2} \sqrt{1-(|\gamma|/2)^2} =M^{-j/2} 
 \sqrt{M^{-j}- M^{-2j}/4} =   O(M^{-j})
\ee
while  $\de\gamma_{\perp}= O(M^{-j/2})$. So this time
we must use rectangles of width $M^{-j}$ in the radial
direction (that of $\gamma_{\Vert}$) and $M^{-j/2}$ in the 
orthogonal direction (see $\ga_3$ in  Fig.\ref{shells}).

\paragraph{4.} Finally when $|\gamma|< M^{-j/2}$ ($\si < M^{-j/2}$) 
we are on the diagonal
terms of the matrix. Here too we have two extreme cases. For
$|\gamma|\simeq M^{-j/2}$ the momentum space should be 
partitioned as above, and there is still only one sector
pair (so only one matrix element $V_{\al\al}$ associated
to the corresponding random variable. But when $|\gamma|\leq M^{-j}$,
then we have only one cube of side $ M^{-j}$ around the value $\gamma=0$
and the corresponding random variable appears in every diagonal term.  

\medskip

In order to interpolate between these extreme situations we have
to introduce shells where $|\gamma|$ is approximately constant. 
In each shell then we introduce cubes or rectangles depending on 
the value of $|\gamma|$. The details are shown below.

\paragraph{Forward-momentum.} This region corresponds to the
annulus between $|\ga_1|$ and $|\ga_2|$ in Fig.\ref{shells} and   
is divided into shells according to an index  $k$, $k=0,..., j/2-1$
so that $M^{-k-1} \le  |p |  \le  M^{-k}$. 

Each forward-shell is now further divided into
rectangular momentum cells of length $O(M^{-j/2})$ in 
the direction parallel to the center of the cell, and of
width $M^{-k-j/2}$ in the direction perpendicular to the
center of the cell.

Finally, for each such cell, the cube $\Delta$ is divided into 
corresponding dual tubes
$\tau$ of length $M^{k+j/2}$ in the direction perpendicular to
the center of the cell, and of width $M^{j/2}$ in the direction
parallel to the center of the cell. Our normalization 
is that each random variable
for a pair $(\ga , \tau)$, which corresponds to a volume 1 in phase space
has covariance 1. These variables are independent Gaussian variables.

\paragraph{Backward momentum region.}

The backward region (region between $|\ga_2|$ and $|\ga_3|$ in 
Fig.\ref{shells}) is
also divided into shells labeled
by a similar index $k=0,..., j/2$, but the backward $k$-shell
corresponds now to momenta 
$2 - M^{-2k} \le  |p |  \le  2- M^{-2k-2}$ for $k<j/2$. The last shell
is simply $2- M^{-j} \le |p| \le 2+ M^{-j}$. Each shell
is now further partitioned
into $O(M^{j-k})$ cells $\ga$ of width $M^{-j/2}$ in 
the direction perpendicular to
the center of the cell, and length $M^{-j/2-k}$ in the direction parallel to
the center of the cell. 
The cube $\Delta$ is again divided into dual tubes $\tau$ 
of width $M^{j/2}$ in the direction perpendicular  to the center of $\ga$
and length $M^{j/2+k}$ in the parallel direction. Each corresponding
random variable $V_{\ga, \tau}$ corresponds again to a different
discretized independent random variable for the potential $V$
with covariance 1 since it corresponds to a volume 1
in phase space.

\paragraph{0-momentum region.}

The last shell is simply $|p| \le M^{-j/2}$. 
The treatment of this region  is subtle because the anisotropic sectors are 
not a basis very well adapted to this zero momentum region.
We remember that the coefficients cannot be totally independent
because in the isotropic basis (treated previously) there is a 
diagonal term proportional to a single variable $V_0 Id$ where
$Id$ is the identity matrix, and this term must be present in any
basis. They are not totally dependent either, so the best 
description consist in
further splitting the  phase-space corresponding to the 0-momentum region
into unit volumes in phase-space. 

Therefore we split the 0-momentum region into shells
$Sh_{0,k'}$, in which the momentum size is
between $M^{-j +k'-1}$ and $M^{-j+k'}$ (for $k'=0$ the lower
bound being simply 0). 
Each such shell is divided into $O(M^{k'})$
cells $\ga'$ of length $O(M^{-j+k'})$ in the direction parallel to the center
of the cell, and  $M^{-j}$ in the perpendicular direction.
The central shell $k'=0$ corresponds to a single cell (the "true"
zero momentum region).

The size of the cells ensures that, given a shell $k'$ and
a matrix element $V_{\al\al}$ only one cell in the shell contributes
to it. On the other hand the same cell may contribute to 
different diagonal terms. In the limit $k'=0$ the same cell appear 
in all diagonal terms.

For each cell we divide the cube into dual tubes $\tau'$, with length $M^j$
in the direction perpendicular to the center of the cell, and width
$M^{j-k'}$ in the parallel direction.
Remark that the total number of pairs $(\ga', \tau')$ which pave
the phase space of this 0-momentum region remains $O(M^j)$
as it should be, and that to each such pair is associated a
random variable of covariance 1.

\medskip

\paragraph{The resulting random matrix.}

Now to model the rhombus rule we have to understand the
random variable $V$ coupling two anisotropic pairs $(\al, t)$ 
and $(\beta, t')$. 
A momentum cell $\ga$ of $V$ defines uniquely four possible
pairs $\alpha, \beta$ (up to nearest neighbors, neglected). Indeed 
we still have Hermiticity and spin flip:
\be V(\al ,\beta) = \bar V(\al ,\beta) = 
V(-\beta ,- \al) = \bar V(-\al ,- \beta) \ .
\ee

Accordingly there are about 
$O(M^k)$ pairs $\al, \beta$ which have their sum in the
forward or backward
shell of index $k$. We consider then two tubes 
$t,t'$ and remark that their intersection is either empty
or roughly of width $M^{j/2}$ and length $M^{j/2 + k}$.
Furthermore as soon as $k$ increases, the direction of 
$\al$ and $\beta$ begins to collapse.
This applies both to forward and backward scattering. 

\begin{figure}
\centerline{\psfig{figure=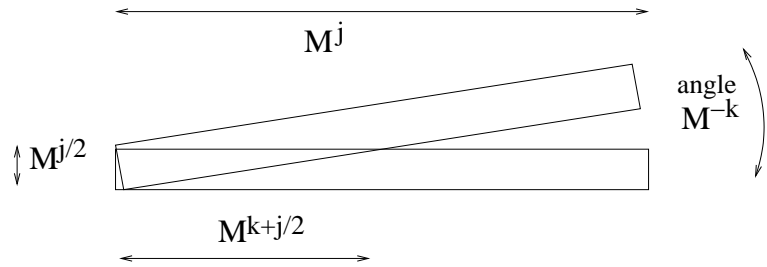,width=10cm}}
\caption{Intersecting tubes} \label{fig5}
\end{figure}

When they intersect, their intersection is a tube of length $M^{k+j/2}$
and width $M^{j/2}$ and there is therefore a single independent variable
$V_{\ga, \tau}$ of covariance $1$ both in the forward scattering case and
in the backward scattering case corresponding to this intersection 
(see Figure\ref{fig5}).

\begin{figure}
\medskip
\medskip
\centerline{\psfig{figure=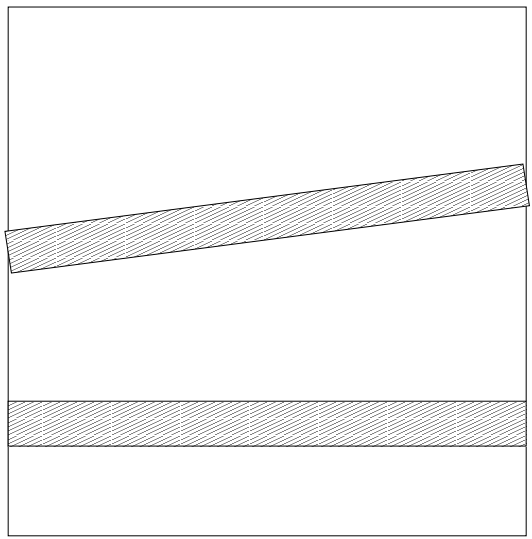,width=7cm}}
\caption{Two missing tubes in a box}\label{fig6}
\end{figure}
\begin{figure}
\centerline{\psfig{figure=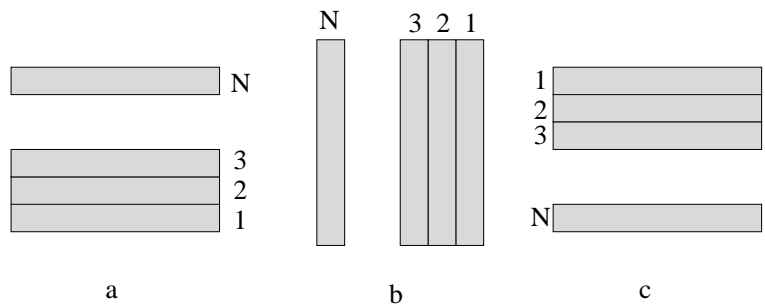,width=10cm}}
\caption{Ordering of the spatial tubes after a rotation of $\pi/2$ (b)
or $\pi$ (c)}
\label{ord}
\end{figure}

When the two tubes miss (see Figure \ref{fig6}), we put a zero 
in the corresponding random matrix.

This describes the random matrix 
if we complete by Hermiticity and symmetry around antidiagonal (spin flip),
except for the diagonal blocks $(\al, t; \al, t')$ which correspond 
to the 0-momentum
region for $V$. The coefficients of these blocks are 0 except on the
pure diagonal $t=t'$, since two distinct parallel tubes $t$ and $t'$
are disjoint.
(Similarly the antidiagonal blocks corresponding to extreme backward scattering
$\al = -\beta$ are purely antidiagonal (0 except on the anti-diagonal). 
But the antidiagonal
is made of independent random variables of covariance 1, while the diagonal
is more complicated).
As we have seen above, a diagonal term is represented as a sum of independent
random variables each corresponding to a different shell $k'$.

Let $(\al, t)$ be a given anisotropic sector and tube, and $k'$
a shell index. We call $\ga' (\al, k')$ the cell of the shell of index
$k'$ perpendicular to the center of $\al$, and we call $\tau' ( t, k')$
the corresponding associated tube containing $t$. The pair
$(\ga'(\al, k'), \tau' ( t, k'))$ corresponds to a single random variable of 
scale $k'$ associated to $(\al, t)$.

The diagonal coefficient of our random matrix is then simply
\be  M_{(\al, t); (\al t)} = 
\sum_{k'=0}^{j/2} V_{\ga'(\al, k'), \tau' ( t, k')} \ .
\ee

Remark that conversely for a given $(\ga' ,\tau')$ in shell $k'$
there are roughly $M^{j/2 -k'}$ tubes $t$ in $\tau'$, and roughly
$M^{j/2 -k'}$ cells $\al$ giving the same $\ga'$, hence 
$M^{j-2k'}$ pairs $(\al, t)$ giving the same $(\ga' ,\tau')$.
This means that a given random variable $V_{\ga', \tau' } $
appears at $M^{j-2k'}$ different diagonal places in the matrix.
For $k'=0$ the single random variable, also called $V_0$
appears everywhere. At the other end, for $k'=j/2$
the variables $V_{\ga', \tau' } $ appear essentially at a single place.

\subsection{The hierarchical approximation}

This approximation consists in 
somewhat simplifying the intersection
rule for $t$ and $t'$ and the angular condition
on $\al$ and $\beta$.
For $(\al \ne \beta)$ instead of considering that the pair is of class
$k$ if their sum lies in the $k$-th shell, we 
divide the set of anisotropic sectors into $M^k$ packets for each value
 of $k$, $k=0,...1,...j/2$
(each packet containing $M^{j/2 - k}$ sectors)
and we replace the rule on the sum $\ga$ to be in shell $k$ by the
simpler "hierarchical" condition
that $\al$ and $\beta$ belong to the same packet of scale $k$ but not to the
same of scale $k+1$.

In the same way we simplify the intersection rule for the tubes. 
We divide for each sector $\al$ the set of tubes $T_{\alpha} $
into $M^k$ packets, according to translation in the 
direction perpendicular to $\al$ and label these packets with an index 
$r_{al}^{(k)}=1,...,M^k$.
Each packet corresponds itself to a larger tube of width $M^{j-k}$, 
hence contains $M^{j/2 -k }$ anisotropic tubes of width $M^{j/2}$.
Let us consider now  two tubes $t, t'$ for a pair 
$(\si_\al\al, \si_\beta\beta)$ 
of category $k$, where $\si_\al=\pm 1$ says if $\al$ is in the
upper half or lower half of the circle. 
Then we slightly simplify the intersection rule by saying that, 
if $\si_\al=\si_\bt$, then   $t, t'$
intersect if their packet index
of degree $k$ is the same: $r_{\al}^{(k)}=r_{\beta}^{(k)}$ 
and miss completely otherwise. On the other hand, when $\si_\al=-\si_\bt$,
for instance $\si_\al=1,\si_\bt=-1$, then the packets in $R_\bt$ have
performed a rotation by an angle $\pi$, and the ordering is reversed
(see Fig.\ref{ord}). In this case we say that  $t, t'$
intersect if their packet index satisfy $r_{\al}^{(k)}=M^k-r_{\beta}^{(k)}$.
 
Finally we take $M=2$, which also simplifies the computation.
In the limit $N\to \infty$ the density of states of the real problem
should be similar to the limit of this hierarchical limit.

The matrix can then be represented as a set of $\sqrt{N}\times\sqrt{N}$
blocks. Each block corresponds to a fixed sector pair, 
and the variables in the block give the couplings between the
tubes. In the  blocks corresponding to $k=1$  
(both in the backward and forward region)  all sectors intersect,
so all elements are independent random variables.
For $k>1$ the block has a block diagonal structure, with 
$2^{j/2-k}\times 2^{j/2-k}$
blocks along the diagonal or the antidiagonal, depending if 
we are considering the forward or the backward region. 
The elements inside the blocks are independent random variables,
the elements outside are zero. In the limit when $k=j/2$ the 
blocks have non zero elements only on the diagonal or antidiagonal.
In the first case, the elements are no longer independent variables.

\begin{figure}
\centerline{\psfig{figure=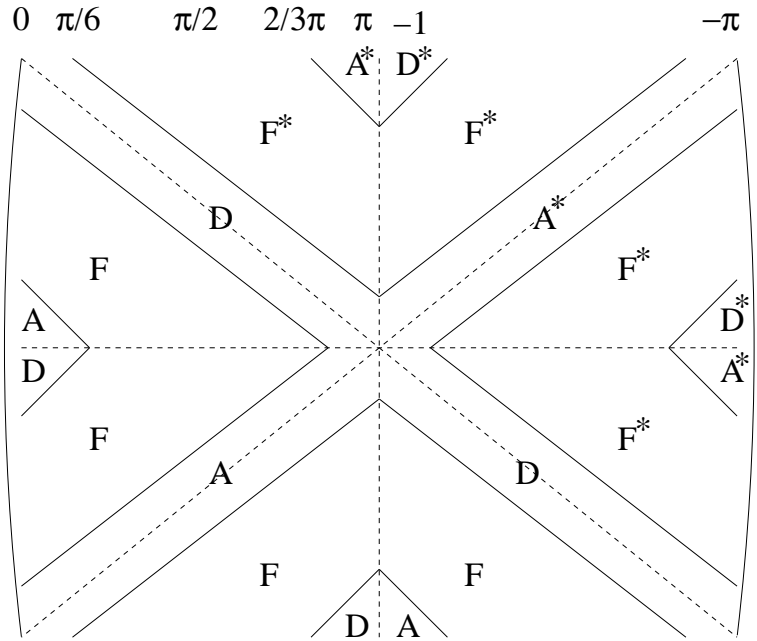,width=10cm}}
\caption{Form of the random matrix after the hierarchical approximation;
the F represent the areas where no zeros appear, D and A are the areas
where only blocks on the diagonal or antidiagonal are non zero. The size of
the block is very big as we are near to the border with the F region and
reduce to one as we go towards the diagonal and antidiagonal}
\label{matr}
\end{figure}

Note that, from the definition of forward and backward shells,
the two $k=0$ shells cover angles $\pi/6< \si < 2/3 \pi$ 
approximately. In this region $k=0$ then all tube packets intersect
and the corresponding block of the matrix has no zeros.
Therefore, for each line in the matrix, at least half of the blocks 
have no zeros. 
For $k>0$ (both in the backward and forward region) we are 
considering only small angles. As a result the matrix looks
like Fig.\ref{matr}.

\begin{figure}
\centerline{\psfig{figure=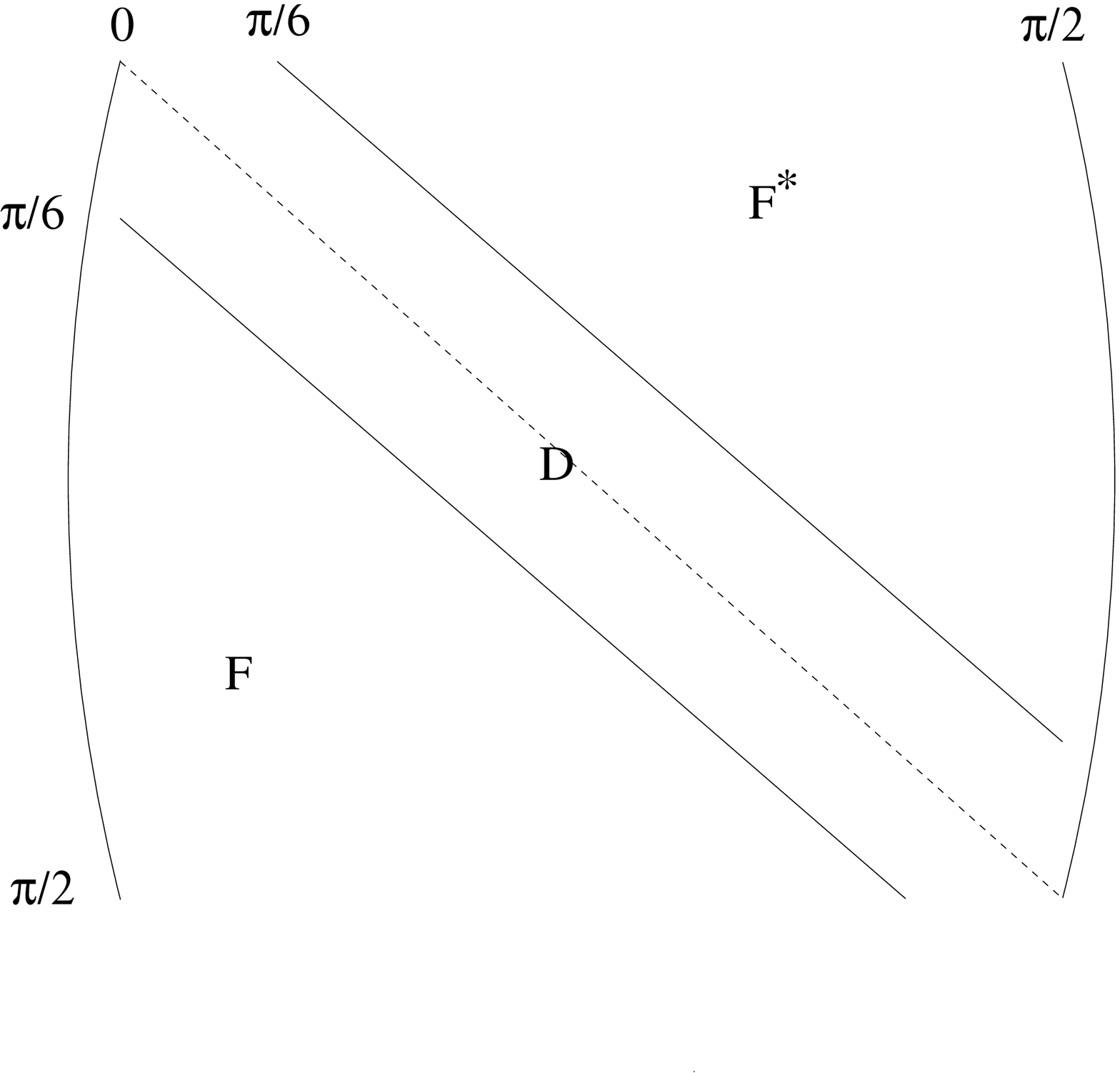,width=5cm}}
\caption{Form of the random matrix if we select only the first $\sqrt{N}/2$
sectors}
\label{redmatr}
\end{figure}

We still have to say what are the diagonal terms $V_{\al,t;\al,t}$.
We may consider only the first $\sqrt{N}$ sectors as the others are
obtained by flip symmetry. The diagonal ${\cal D}$ can then be written
as a sum of diagonal matrices
\be
{\cal D} = \sum_{k'=0}^{j/2} {\cal D}_{k'} \ .
\ee
For $k'\neq k''$ the corresponding random variables are independent.
For a fixed $k'$ ${\cal D}_{k'}$ is again a diagonal matrix, made
of $2^{k'}$  diagonal blocks $B(k')_r$ $r=1,...2^{k'}$.
Each block $B(k')_r$ actually selects a subset of $2^{j/2-k'}$
sectors that are associated to the same set of random variables. 
For $k'=0$ there is only one block as  
${\cal D}_{0}=V_0 Id$. In the opposite case $k'=j/2$ there are 
$2^{k'}$ as all diagonal elements  are independent
random variables.
As an example, for $k'=1$ we have only two blocks
\be
 {\cal D}_{1} = 
\left (
\ba{cc}
B(1)_1 & 0   \\
0 &   B(1)_2 \\       
\ea
\right ) \ .
\ee

Then a single block $B(k')_r$ is built from  $2^{j/2-k'}$ identical
sub-blocks $V(k',r)$, one for each sector. This sub-block  $V(k',r)$
actually identifies the set of independent random variables associated
to this set of sectors.
For $k'=0$ the single block $B(0)$ has   $2^{j/2}$ sub-blocks, that is
all sectors are associated to the same set of variables. 

Finally, we have to write  the $\sqrt{N}\times \sqrt{N}$  
block $V(k',r)$ which corresponds
to the tube indices. This block contains $2^{k'}$ independent random
variables. In order to distribute them on the diagonal 
we introduce a partition of the set $\{1,...,\sqrt{N}\}$
in $2^{k'}$ sets of $2^{j/2-k'}$ indices, defined by $I=\cup_{l=1}^{2^{k'}} I_l$ and
\be
I_l = \{i \ |\ 2^{j/2-k'} (l-1) < i \leq  2^{j/2-k'} l\}\ .
\ee
Now to each $l=1,...,2^{k'}$ we associate an independent random variable 
$ V(k',r)_l$.
With this notations
\be
V_{ii} = V(k',r_{k'})_{l_k'} \qquad {\rm if}\ i\in I_l\ .
\ee
Again we see that in the case $k'=0$ we have only one set $I$ so that
there is only one independent random variable.

\subsubsection{Toy model}
To understand how the analysis is going to change with respect
to the isotropic model we first consider a simplified model,
where we restrict to the sectors $1,...\sqrt{N}/2$, that is
to angles $0\leq \si\leq \pi/2$. 
The matrix then looks like Fig\ref{redmatr}.

We can construct now this matrix in a hierarchical way.

The first hierarchical step is $k=1$, with two packets of sectors, and
two packets of tubes. If we list the indices of the first packet first and
second packet second, and then within each sector
packet we list the tube indices with first packet of tubes first, this
cuts the matrix into 16 blocks. Then we put random independent
coefficient (with the Hermiticity condition) everywhere
except in the two upper blocks $B_{12}$ and $B_{34}$
which we fill with zeros. (We complete by Hermiticity
so that lower blocks $B_{21}$ and $B_{43}$ are also zero).
These zeros correspond to the first group of missing tubes.
In this way we get an $N \times N$ called matrix $M_1(N)$

\be
M_1(N) = \left (
\ba{cccc}
B_{11} & 0 & B_{13} & B_{14}  \\
0 & B_{22}     & B_{23} & B_{24}  \\
B_{13}^* &B_{23}^*  & B_{33} & 0  \\
B_{14}^* &B_{24}^*  & 0 &  B_{44} \\
\ea
\right ) =
\left (
\ba{c|c}
\ba{cc}
B & 0 \\
0 & B \\ \ea
& B^*  \\
\hline
B & 
\ba{cc}
B & 0 \\
0 & B \\ \ea   \\
\ea
\right )
\ee
where $B_{ii}^*=B_{ii}$ for all $i=1,..,4$ and each block 
$B_{ij}$ is a  $N/4 \times N/4$ matrix.

Then for $k=2$, the 4 previous diagonal blocks 
of $M_1(N)$ are themselves
replaced by $M_1 (N/4)$. This defines $M_2(N)$, which is
a sixteen by sixteen matrix.

{\small\be
\left (
\ba{c|c} \ba{c|c} \ba{c|c} \ba{cc}
\hspace{-.2cm} B &  0 \hspace{-.2cm}  \\
\hspace{-.2cm}  0 & B  \hspace{-.2cm} \\ 
\ea
& B^*  \\
\hline
B & 
\ba{cc}
\hspace{-.2cm}B & 0 \hspace{-.2cm}\\
\hspace{-.2cm}0 & B \hspace{-.2cm}\\ \ea   \\
\ea
&  0 \\
\hline
0 &  
\ba{c|c} \ba{cc}
\hspace{-.2cm}B & 0\hspace{-.2cm} \\
\hspace{-.2cm}0 & B\hspace{-.2cm} \\ \ea
& B^*  \\
\hline
B & 
\ba{cc}
\hspace{-.2cm}B & 0\hspace{-.2cm} \\
\hspace{-.2cm}0 & B\hspace{-.2cm} \\ \ea   \\
\ea 
\\
\ea
&   B^* \\
\hline
B  & 
\ba{c|c}  \ba{c|c} \ba{cc}
\hspace{-.2cm}B & 0\hspace{-.2cm} \\
\hspace{-.2cm}0 & B\hspace{-.2cm} \\ \ea
& B^*  \\
\hline
B & 
\ba{cc}
\hspace{-.2cm}B & 0\hspace{-.2cm} \\
\hspace{-.2cm}0 & B\hspace{-.2cm} \\ \ea   \\
\ea
&  0 \\
\hline
0 &  
\ba{c|c} \ba{cc}
\hspace{-.2cm}B & 0\hspace{-.2cm} \\
\hspace{-.2cm}0 & B\hspace{-.2cm} \\ \ea
& B^*  \\
\hline
B & 
\ba{cc}
\hspace{-.2cm}B & 0\hspace{-.2cm} \\
\hspace{-.2cm}0 & B\hspace{-.2cm} \\ 
\ea \\ \ea\\ \ea \\ \ea\right )
\nonumber\ee} 

We iterate. To get $M_k(N)$ we replace each of the $4^{k-1}$ diagonal blocks
 of $M_{k-1}(N)$ by $M_1(N/4^{k-1})$. We stop at $k=j/2$. Then
the diagonals blocks of size 1 have been reached.

Finally in the last step we change the diagonal coefficients
of $M_{j/2}(N)$. We replace each of them by the appropriate
sum of independent variables according to the analysis of the
zero momentum region. Again this is done in a hierarchical way. 
At $k=0$ we put a $V_0$ at every diagonal point. 
At $k=1$ we split the diagonal in two parts. To all terms $V_{ii}$
with $i\leq N/2$ we add the variable $V_1^1$, to all others
$V_1^2$. To iterate that we assign an ordering $r_k(i)$. 
For $k=1$ we have $r_k(i)=1$ for all $i\leq N/2$ and  $r_k(i)=2$
otherwise. At $k=2$ we split in four sets and iterate.

The diagonal coefficient $V_{i,i}$ of the final matrix is therefore

\begin{equation}
V_{ii} = V_0 + ... + V_k^{r_k(i)} + V_{j/2} ^{r_{j/2}(i)}
\end{equation} 
where at each scale $k$ there is $4^{k-1}$ variables $V_k^r$, 
each serving for $N/4^{k-1}$ coefficients,
and $r_k(i)=1,... ,\; 4^{k-1}$ is simply the label for  the variable of the cell 
in the shell $k$ associated to  $i$.

The rest of the matrix remains 
unmodified, with independent variables.

So the matrix is made of

- independent random variables at each non-diagonal place, except where
the zeros of $M_{j/2}(N)$, lie, which correspond to tubes missing each other,

- a more complicated sum of random variables at each diagonal place.
Remark in particular that this sum always contains the $V_0$
variables, common to all the diagonal.

Again the density of states for this matrix
should follow some semi-ellipse Wigner's law when $N \to \infty$.
We have to reproduce the computation of the previous section
but with additional fields which complicate the picture.
Each set of zeros should indeed be represented 
by a new pair of complex fields which suppress
the off diagonal blocks with zeros in the quartic form after
integrations of $V$.

When searching for a main saddle point, an iterative solution seems doable
by plugging the smaller blocks solutions into large means fields,
until the last one for which we should get
$\nu (E) =    C_k \sqrt {E_k^2 - E^2}$ where $E_k$ lies between
$\sqrt 2$ and $\sqrt 3$ \cite{DR}.

\subsection{A Hierarchical 
Isotropic matrix with correct Momentum conservation rule}

An alternative is to stick to the initial basis of isotropic sectors but
to take seriously into account the necessary modification of the rhombus
rules for degenerate rhombuses.

The approximation in which each random variable corresponds
roughly to a single pair of momenta, up to Hermiticity and flip
is indeed not correct. When the momentum of $V$
is of size $M^{-k}$ (forward) or $2- M^{-2k}$ (backward),  
there are roughly $M^{k}$ pairs that have
this transfer momentum.

Taking again $M=2$ the hierarchical approximation to this correct conservation
rule consists in identifying blocks of $2^k$ by $2^k$ coefficients to a single
random variable as approaching the diagonal (or anti-diagonal
in the version with the flip symmetry).

In this point of view the hierarchy is indexed by $k$ from 0 to $j$ 
(not $j/2$).
This model should again give the same limit as the previous one when
$j \to\infty$.

\section{The isotropic 3D Model}

We would like to conclude with a remark concerning the three dimensional model.
The $U(1)$ symmetry which generalizes the flip symmetry in that 
case has orbits which are
quite complicated to describe because there is no simple labeling 
of sectors on the sphere
which respect the rotation invariance of the sphere. For instance 
labeling according to spherical coordinates introduces a preferred polar 
axis. So here again it may be instructive
to mimic this $U(1)$ symmetry by treating first some hierarchical 
approximations in which
the $U(1)$ symmetry is replaced by a discrete $Z_p$ symmetry, with $p=2^k$.
For instance such symmetries could act in a simple way on coefficients, by 
folding
$k$ times the upper triangle around its bisecting line, and identifying the 
$2^k$ coefficients
at the same position in all the layers. Then letting $p \sim \sqrt N$
and $N\to \infty$, it should be 
possible to prove again through a supersymmetric analysis
that the density of states of such a model tends to Wigner's law. 

Such a result would certainly
increase our confidence that the density of states for the true
three dimensional Anderson model falls into the same universality class
than the GUE and obeys Wigner's semi-circle law.


\begin{thebibliography}{99}

\bibitem{BMR}
{J. Bellissard, Jacques Magnen, V. Rivasseau,  
Markov Processes and Related Fields, 9, 1-30 (2003)}

\bibitem{Voi} {D. Voiculescu, Limit Laws for Random Matrices and Free Products,
Invent. Math. {\bf 104}, 201 (1991)}

\bibitem{DPS}
{M.~Disertori, H.~Pinson and T.~Spencer, Density of states for random 
band matrix,
Commun. Math. Phys,  {\bf 232}, 83 (2002)}.
\bibitem{Mehta} {M. L. Mehta, Random Matrices, Academic Press, 1991}

\bibitem{Past}
{L.~Pastur and A.~Figotin, Spectra of Random and Almost-Periodic
Operators,
Springer-Verlag 1982}

\bibitem{Mir}
{A.~Mirlin,
Statistics of energy levels and eigenfunctions in
disordered and chaotic systems: supersymmetry approach.
{\it Phys. Rep.} {\bf 326}, 259--382 (200)}

\bibitem{Zirn} {M. Zirnbauer, Journ. Math. Phys. {\bf 37}, 4986 (1996)}

\bibitem{Poir}
{G.~Poirot, These, Ecole Polytechnique, 1998}


\bibitem{EFSHK}
{B.I.~Shklovskii and A.L.~Efros,
{\it Electronic Properties of Doped
Semiconductors}, Springer (1984)}

\bibitem{And58}
{P.W.~Anderson,
Absence of diffusion in certain random lattices
{\it Phys. Rev.\/} {\bf 109}, 1492--1505, (1958)}

\bibitem{gang4}
{E.~Abrahams, P.W.~Anderson, D.C.~Licciardello and
T.V.~Ramakrishnan,
Scaling theory of localization: absence of quantum
diffusion in two dimensions.
{\it Phys. Rev. Lett.} {\bf 42}, 673--676, (1979)}


\bibitem{GMP}
{I.~Gold'sheid, S.~Molchanov and L.~Pastur,
A random homogeneous Schr{\"o}dinger operator has
a pure point spectrum,
 {\it Funct. Anal. Appl.} {\bf 11}, 1--10 (1977)}

\bibitem{KuSo}
{H.~Kunz and B.~Souillard,
Sur le spectre des op{\'e}rateurs aux
diff{\'e}rences finies al{\'e}atoires.
{\it Commun. Math. Phys.} {\bf 78}, 201--246 (1980)}

\bibitem{FMSS}
{J.~Fr{\"o}hlich, F.~Martinelli, E.~Scoppola and
T.~Spencer,
Constructive proof of localization in the Anderson
tight binding model.
{\it Commun. Math. Phys.} {\bf 101}, 21--46 (1985)}

\bibitem{AiMo}
{M.~Aizenman and S.~Molchanov,
Localization at large disorder and at extreme
energies: an elementary derivation,
{\it Commun. Math. Phys.} {\bf 157}(2), 245--278 (1993)}

\bibitem{Bergmann}
{G.~Bergmann, Weak localization in thin films, a time-of-flight
experiment with conduction.
{\it Phys. Rep.} {\bf 107}, 1--58 (1984)}

\bibitem{Kra}
{B.~Kramer and A.~MacKinnon,
Localization: theory and experiments.
{\it Rep. Prog. Phys.} {\bf 56}, 1469--1564 (1993)}


\bibitem{Efet83}
{K.B.~Efetov,
Supersymmetry and theory of disordered metals.
{\it Adv. Phys.} {\bf 32}, 53--127.
(1983)}

\bibitem{Kra2}
{I.Kh.~Zharekeshev, M.~Batsch and B.~Kramer,
Crossover of level statistics between strong and
weak localization in two dimensions.
{\it Europhys. Lett.} {\bf 34}, 587--592, (1996)}

\bibitem{AltShk}
{B.~Altshuler and B.I.~Shklovskii,
Repulsion of energy levels and conductivity of
small metal samples.
{\it Sov. Phys. JETP\/} {\bf 64}, 127--135 (1986)}

\bibitem{Efet}
{K.B.~Efetov, {\it Supersymmetry in Disorder and Chaos.}
Cambridge University Press, 1997}

\bibitem{CHKR} {Jean-Michel Combes, Peter D. Hislop, Fr\'ed\'eric Klopp and Georgi Raikov,
Global continuity of the integrated density of states for random Landau Hamiltonians,
Proceedings of the Marseille conference in honor of L. Pastur, 2002}

\bibitem{BG}
{G.~Benfatto and G.~Gallavotti, Perturbation theory of the Fermi surface in a
quantum  liquid, a general quasi-particle formalism and one
dimensional systems.
{\it J. Stat. Phys.} {\bf 59}, 541--664 (1990)}

\bibitem{FT}
{J.~Feldman and E.~Trubowitz,
Perturbation theory for many fermions systems.
{\it Helv. Phys. Acta} {\bf 63}, 156--260 (1990)}

\bibitem{FMRT}
{J.~Feldman, J.~Magnen, V.~Rivasseau and
E.~Trubowitz,
An infinite volume expansion for many fermion
green's functions.
{\it Helv. Phys. Acta} {\bf 65}, 679--721 (1992)}

\bibitem{MPR1}
{J.~Magnen, G.~Poirot and V.~Rivasseau,
The Anderson model as a matrix model,
{\it Nucl. Phys. B} {\bf 58}, 149--162, (1997)}

\bibitem{constr}
{Constructive Physics, 
Lecture Notes in Physics 446, Springer Verlag (1995)}

\bibitem{R} {V. Rivasseau,  The two dimensional Hubbard Model at half-filling:
I. Convergent Contributions, Journ. Stat. Phys. 106(3): 693-722; (2002)}

\bibitem{Po}
{G.~Poirot,
Mean Green's function of the Anderson model at
weak disorder with an infrared cutoff. 
{\it Ann. Inst. H.~Poincar\'e} {\bf 70},
101--146, (1999)}


\bibitem{MPR2}
{J.~Magnen, G.~Poirot and V.~Rivasseau
Ward type identities for the 2d Anderson model
at weak disorder, J. Stat. Phys. {\bf 93}, 331--358, 1998}


\bibitem{DR}
{M. Disertori and V. Rivasseau, in preparation}





\end{thebibliography}
\end{document}